\documentclass[preprint,aps,amsmath]{revtex4}
\usepackage{graphicx}
\usepackage{amssymb}
\usepackage{bm}
\usepackage{subcaption}

\newcommand\bnabla{\boldsymbol \nabla}
\newcommand\bs{\boldsymbol}

\begin{document}

\title{TIFF: Gyrofluid Turbulence in Full-f and Full-k} 
\author{Alexander Kendl}
\affiliation{Institut f\"ur Ionenphysik und Angewandte Physik, Universit\"at
  Innsbruck, Technikerstrasse 25, 6020 Innsbruck, Austria  \vspace{2cm}
} 
\email{alexander.kendl@uibk.ac.at}

\begin{abstract}

   \vspace{1cm}

A model and code (``TIFF'') for isothermal gyrofluid computation of
quasi-two-dimensional interchange and drift wave turbulence in magnetized
plasmas with arbitrary fluctuation amplitudes (``full-f'') and arbitrary 
polarization wavelengths (``full-k'') is introduced.
The model reduces to the paradigmatic Hasegawa-Wakatani model in the limits of
small turbulence amplitudes (``delta-f''), cold ions (without finite Larmor
radius effects), and homogeneous magnetic field.
Several solvers are compared for the generalized Poisson problem, that is
intrinsic to the full-f gyrofluid (and gyrokinetic) polarization equation, and
a novel implementation based on a dynamically corrected Fourier method is
proposed.  
The code serves as a reference case for further development of
three-dimensional full-f full-k models and solvers, and for fundamental
exploration of large amplitude turbulence in the edge of magnetized plasmas.
\end{abstract}

\maketitle

\section{Introduction: gyrofluid models and polarization}

Turbulence in magnetized plasmas, which is generally driven by ubiquitous
gradients of density and temperature, is a subject of considerable interest
and importance in fusion energy research with toroidal magnetic
high-temperature plasma confinement experiments such as tokamaks and
stellarators \cite{Tynan09}. 
Models for instabilities and turbulence in magnetized plasmas are based on
gyrokinetic, gyrofluid or fluid drift descriptions \cite{Scott21a,Scott21b}. 

Gyrofluid theory as a fluid-like model for the low-frequency dynamics of
gyrocenter densities in magnetized plasmas including finite Larmor radius
(FLR) effects has been pioneered in the late 1980s \cite{Knorr88}. 
In the early 1990s collisionless Landau closure has been applied to derive
three-dimensional  gyrofluid models \cite{Hammett90} for computations in slab
\cite{Dorland93} and  toroidal geometry \cite{Beer96}.
The approach was further generalized to electromagnetic
models \cite{Snyder01,Scott00}.
Energetically consistent 6-moment electromagnetic gyrofluid equations with FLR
effects have subsequently been systematically derived from the corresponding gyrokinetic
theory \cite{Scott05,Kendl10}. 
 
Gyrofluid theory has its place in the hierarchy of magnetized plasma models
somewhere in between gyrokinetic and drift-reduced fluid theory
\cite{Scott00,Scott05,Scott06,Scott07,Scott10,Scott10CPP,Kendl10}.
The quality of a model of course not only depends on the rank within some
straight hierarchy, but also on multiple secondary modelling assumptions
that are introduced in any practical application and numerical simulation.
Gyrofluids have the advantage over fluid theories to consistently model
some kinetic effects such as finite Larmor radii or Landau damping.
They do not depend on (Braginskii type) collisional closures, but collisional
effects can be amended. Compared to gyrokinetic modelling of
a 5-dimensional distribution function, the 3-dimensional gyrofluid moment
models can be computationally much more efficient.

In ``full-f'' (a.k.a.~``total-f'') gyrofluid models
\cite{Strintzi04,Strintzi05,Madsen13}, similar to full-f gyrokinetic models
that evolve the total distribution function, no assumptions on the smallness of
fluctuation  amplitudes $f({\bf x},t) = f_0({\bf x}) + \tilde f ({\bf x},t)$
are made, instead of evolving only a small ``delta-f'' deviation $\tilde f$
from a static background equilibrium $f_0$ \cite{Lee83,Hahm88,Sugama00,Brizard07,Scott10GK,Krommes12,Scott16}. 
In fusion plasmas this is particularly relevant for the edge and scrape-off
layer (SOL) regions, where large-amplitude blob or edge localized mode (ELM)
filaments dominate the transport \cite{Endler95,Zweben07}. 

The development of full-f gyrokinetic codes started a bit more than a decade ago (see, for example,
refs.~\cite{Heikkinen08,Grandgirard07,Ku09,Dorf16,Grandgirard16,Idomura08,Scott10CPP,Idomura14,Pan18,Shi19,Scott16,michels21}),
and often involves long-term development in large research groups or
collaborations. All these (perpetually evolving) codes employ different levels
of secondary modelling assumptions and approximations, such as for geometry
(full torus vs. flux tube; circular vs. X-point; core vs. edge/SOL), for
polarization (linear vs. nonlinear), or for collision operators,
electromagnetic effects, and (SOL and core) boundary conditions. 

Full-f gyrofluid models are directly related to gyrokinetics, as they are usually obtained by taking
appropriate fluid moments of the full-f gyrokinetic equations
\cite{Madsen13}. Therefore the difference between delta-f and full-f gyrofluid
models involves the same choice of assumptions on the level of the gyrokinetic
Lagrangian action, which in principle requires a consistent decision between
either small perturbations but applicability for arbitary wave lengths, or
arbitrary perturbation and flow amplitudes but restriction to long wave lengths
\cite{Scott10,Strintzi04,Strintzi05}.

For simulations of full-f gyrofluid models, over the last years two separate
code implementations were developed
alongside, which are both based on the same or similar model sets 
\cite{Madsen13,Held17Thesis}, but use largely different numerical approaches:
The modular open source code suite FELTOR (``Full-F ELectromagnetic code in TORoidal geometry'')
\cite{Wiesenberger19,FeltorV6} has been primarily developed and maintained by
Wiesenberger and Held {\sl et al.} \cite{Wiesenberger14,Held16,Held18,Held19};
and the code family TOEFL is being primarily developed by Kendl and
includes the 2d code branch ``TIFF'', which is reported herein.
The acronym TOEFL denotes ``TOkamak Edge FLuid'', and TIFF is ``TOEFL In
Full-f and Full-k''. In its first (drift-Alfv\'en fluid in C/C++)
implementation, TOEFL was largely tantamount to DALF \cite{Scott02} but in
another language. 

The 3d delta-f gyrofluid version T3P in the TOEFL set had been designed to be
comparable with the GEM3 model and code by Scott \cite{Scott03} and Ribeiro
\cite{Ribeiro08}, and was applied in
refs.~\cite{Kendl14,Meyer16,Meyer17,Meyer17b}, and as 2d delta-f gyrofluid
model reduction in refs.~\cite{Kendl12,Meyer17,Kendl18}. 
Full-f low-k 2d and 3d gyrofluid versions of TOEFL in the previous long wavelength
approximation implementation, with otherwise similar numerics as employed here for TIFF, have
already been applied in refs.~\cite{Kendl15,Kendl17,Kendl18b}.
These previous delta-f and full-f low-k simulations can serve as test cases
for the implementation of the present full-f full-k model in its respective parameter limits.
The dual FELTOR vs.~TOEFL code development strategy allows cross-verification,
but they for example also have optimized applicability for different
geometries. Whereas FELTOR can be applied to full global 3d torus geometry,
TOEFL is presently designed for locally field-aligned 3d flux-tube type
toroidal simulation geometry.

What ever gyrofluid moment sets (such as thermal or isothermal, 2d or 3d)
are treated in the codes, all gyrocenter densities are evolved and coupled to obtain the
electric potential $\phi({\bf x},t)$ via the polarization equation, which is
isomorphic to its full-f gyrokinetic equivalent, and is nothing but
the gyrocenter density ($N_s({\bf x},t)$) formulation of quasi-neutrality,
summing up all species (index $s$) charge and polarization densities. 

Until recently, both full-f gyrofluid code families have treated the usual, consistent
long-wavelength form of the full-f polarization equation \cite{Strintzi04,Strintzi05,Madsen13}:
\begin{equation}
  \sum_s \;\left[ \bnabla \cdot \left( \frac{ m_s N_s({\bf x}) }{ B^2}
      \bnabla_{\perp} \phi({\bf x}) \right)  +  q_s \; \text{G}_{1s} N_s ({\bf x}) \right]
  = 0.
\label{eq:pol-lwl}
\end{equation}

This can be re-cast into a generalized 2d Poisson equation:
\begin{equation}
  \bnabla \cdot \varepsilon \; \bnabla_{\perp} \phi = \sigma 
\label{eq:pol-lwl-general}
\end{equation}
where $ \varepsilon ({\bf x}) \equiv \sum_s {m_s N_s ({\bf x}) / B^2}$, and
$\sigma ({\bf x}) \equiv -  \sum_s q_s \; \text{G}_{1s} N_s ({\bf x})$.
Here $m_s$ is the mass and $q_s = Z_s e$ the charge of plasma species $s$ with
gyrocenter density $N_s$. The plasma species include electrons and at least one
but often several ion species. In this present work only one main ion species
(index $s=i$) in addition to the electrons (index $s=e$) is considered.
The full-f gyrofluid generalization to multiple ion species has been discussed
in refs.~\cite{Kendl17,Kendl18b,Reiter23}. 

In the case of a pure $e$-$i$ plasma, the polarization of the electrons can be
neglected because $m_e \ll m_i$, so that  $ \varepsilon ({\bf x}) \approx  {m_i N_i ({\bf x}) / B^2}$.
The gyrocenter densities $N_s$ in $\sigma$ are affected by gyro-averaging, which is
denoted by the operator $G_{1s}$. This can be expressed as $\text{G}_1(b_s) =
\text{G}_0^{1/2}(b_s)$ in wavenumber space with the gyro-screening operator
$\text{G}_{0s} \equiv I_0(b_s) e^{-b_s}$ for $b_s =(\rho_s k_{\perp})^2$ \cite{Scott10}.
This form (derived from velocity integration of the gyrokinetic pendant
including Bessel functions) makes use of the modified Bessel function of the
first kind $I_0$. The Larmor radius $\rho_s = \sqrt{T_s m_s}/(eB)$ of the
particle species ($s$), with temperature $T_s$ and mass $m_s$, normalizes the
$k_x$ and $k_y$ wavenumber components in the 2d plane locally perpendicular to the
magnetic field ${\bs B} = B {\bf e}_z$. The gyro-averaging operators can be
efficiently approximated \cite{Scott10} by their Pad\'e forms $G_{0}(b_s)
\approx \Gamma_{0}(b_s) \equiv 1/(1+b_s)$ and $G_{1}(b_s) \approx \Gamma_{1}(b_s) \equiv 1/(1+b_s/2)$.

The full dynamical nonlinearity $\bnabla \cdot \varepsilon(x,y,t) \; \bnabla_{\perp}
\phi$  in the long-wavelength polarization equation is here retained. 
Most full-f gyrokinetic implementations so far approximate this term by using either only radial
variations of the static background $\varepsilon_0(x)$, or completely linearise it to $\sim \varepsilon_0
\bnabla_{\perp}^2 \phi$ and so neglect spatial and temporal variations in the polarization.

However, the general delta-f form of the polarization equation
\begin{equation}
  \sum_s \;\left[ \frac{ q_s e N_0 }{T_s} \left( \Gamma_{0s} -1 \right) \phi  +
    q_s N_0 \; \Gamma_{1s} \frac{ \tilde{N}_s }{ N_0 } \right]  = 0,
\label{eq:pol-df}
\end{equation}
for small perturbations $\tilde N_s$ on a reference background density $N_0$,
differs from this linearised long-wave length polarization by an additional FLR
contribution proportional to the Larmor radius $\rho_s$, and only agrees in lowest order
Taylor expansion in $b = (\rho_s k_{\perp})^2$ \cite{Scott10}:
\begin{eqnarray}
   \frac{ e^2 Z_s N_0 }{T_s} \left( \Gamma_{0s} -1 \right) \phi & = &   \frac{m_s
     Z_s N_0}{B^2} \frac{\bnabla_{\perp}^2}{1-\rho_s^2 \bnabla_{\perp}^2} \phi
  \\ \nonumber
   & \neq &   \frac{m_s Z_s N_0}{B^2} \bnabla_{\perp}^2 \phi
     \equiv   \varepsilon_0 \bnabla_{\perp}^2 \phi
\label{eq:pol-difference}
\end{eqnarray}

In any case consistency throughout the equations has to be ensured, for example in full-f models
by keeping E-cross-B energy terms in the generalised potential \cite{Scott10,Madsen13}.
Computation of delta-f gyrofluid turbulence with the exact model in comparison
to the delta-f long-wavelength approximation such as in
eq.~(\ref{eq:pol-difference}) show that the differences can be rather pronounced.
This has motivated the development of a consistent arbitrary wavelength
full-f gyrofluid polarization model \cite{Held20}. 
A first implementation of this ``full-f full-k'' model in the code FELTOR and
application to simulation of interchange driven ``blob'' perturbations in a
magnetized plasma is presented by Held and Wiesenberger in ref.~\cite{Held23}. The results therein
clearly show the relevance of arbitrary wave length polarization for
interchange drift modes.

In the present work, an independent code implementation (TIFF) of the ``full-f
full-k'' model and application to 2d drift instabilities and turbulence is given.
In the respective interchange blob mode limit, the results are cross-verified
with the recent FELTOR code results. Different solvers for the underlying
generalized Poisson problem are  compared. An efficient solver based on a
dynamically corrected Fourier method is proposed and tested.

\section{Full-f full-k 2d gyrofluid turbulence model}

The arbitrary wave length full-f model derived in ref.~\cite{Held20} and first
applied by Held and Wiesenberger in ref.~\cite{Held23}, is in the following
implemented in a form which also includes the full-f formulation of the
gyrofluid generalization \cite{Held18,Held19} of the quasi-2d (modified)
Hasegawa-Wakatani (HW) drift wave turbulence model \cite{Hasegawa83,Numata07}.

\subsection{Gyrocenter density equations}

The set of isothermal full-f full-k gyrofluid equations is based on dynamical
evolution equations of the gyrocenter densities for each species $s$ 
in the general form of a continuity equation:
\begin{equation}
  \partial_t N_s  + \bnabla \cdot ( N_s {\bs U}_s ) = 0.
  \label{eq:continuity}
\end{equation}

The velocities ${\bs U}_s = {\bs U}_E + {\bs U}_B + {\bs U}_{\parallel}$ include
the gyro-center E-cross-B drifts ${\bs U}_E = (1/B^2) {\bs B} \times
\bnabla_{\perp} \phi_s$, the gradient-B drifts ${\bs U}_B = (T_s/q_s B^2) {\bs B}
\times \bnabla_{\perp} \ln (B/B_0)$, and parallel velocities  ${\bs   U}_{\parallel}$. 
In contrast to the corresponding fluid continuity equation for particle
densities, the gyrofluid formulation for gyrocenter densities does not contain
polarization drifts, whose effects are covered by the relation of the gyrocenter
densities within the polarization equation for the electric potential $\phi$.
The resulting set of equations can be seen as a variation of the
vorticity-streamfunction formulation for a 2D Euler fluid model. The electric
potential here takes the role of a streamfunction for the advecting E-cross-B
velocity. 

The gyrofluid potentials $\phi_s = \Gamma_{1s} \phi + \Psi_s$ in the E-cross-B
drift ${\bs U}_E$ include the gyro-average part, and the consistent full-f full-k form \cite{Held20} of the
polarization contribution through the E-cross-B energy as $\Psi_s = (m/2qB^2)
|\nabla_{\perp} \sqrt{\Gamma_0} \phi|^2$. For electrons $\phi_e \approx \phi$
can be used because of the small mass ratio and associated small Larmor radius
in comparison to ions. The potential $\phi$ is retrieved from solution of the polarization equation.

\subsection{Polarization equation}

The consistent arbitrary wavelength full-f polarisation equation has been
derived in ref.~\cite{Held20} and is used  here in isothermal
(constant gyroradius) form as in ref.~\cite{Held23}:

\begin{equation}
\sum_s q_s N_s - \bnabla \cdot ( {\bs P}_1 + {\bs P}_2 ) = 0,
  \label{eq:pol-fk}
\end{equation}
where the polarization densities are given as
\begin{eqnarray}
  {\bs P}_1 &=& - (1/2) \sum_s q_s \bnabla_{\perp} \Gamma_1 \rho_s^2 N_s , \\
  {\bs P}_2  &=& - \sum_s \left( \sqrt{\Gamma_0} \frac{q_s N_s}{\Omega_s B}
  \sqrt{\Gamma_0} \bnabla_{\perp} \phi \right).
  \label{eq:poldens}
\end{eqnarray}
Here, in the (isothermal) constant gyroradius approximation the $\Gamma_1$ and
$\sqrt{\Gamma_0}$ operators are self-adjoint, and may for example be evaluated efficiently
in ${\bs k}$ space. The general form  for $\rho_s = \rho_s({\bs x})$ is given in ref.~\cite{Held20}.

For the present implementation in the code TIFF only one ion species
(e.g. Deuterium) is treated, and the electron polarization is neglected.
The full-f full-k polarization equation then can be re-written as:
\begin{equation}
  \bnabla \cdot \sqrt{\Gamma_0} \varepsilon_i \sqrt{\Gamma_0}
  \bnabla_{\perp} \phi  = \sigma
\label{eq:pol-tiff}
\end{equation}
where $\sigma ({\bf x}) \equiv -  \sum_s  Z_s \Gamma_{1s} N_s ({\bf x})$,
with $\Gamma_{0}$ (here for ions only) and $\Gamma_{1s}$ given in second order
accurate Pad\'e approximation \cite{Held20,Held23}, as given above, and
$\varepsilon_i ({\bf x})= {m_i N_i ({\bf x}) / B^2}$. 
In the constant gyroradius (isothermal) case the $\sqrt{\Gamma_0}$ operators
commute with the $\bnabla$ operators, so that also $ \sqrt{\Gamma_0} \bnabla
\cdot \varepsilon_i \bnabla_{\perp} \sqrt{\Gamma_0} \phi  = \sigma$ holds.
By defining $\phi_G \equiv \sqrt{\Gamma_0} \phi$ and $\sigma_G \equiv
\sqrt{\Gamma_0}^{-1} \sigma$, the arbitrary wave length polarization
eq.~(\ref{eq:pol-tiff}) can again be re-cast into the usual form of a
generalised 2d Poisson equation as $\bnabla \cdot  \varepsilon_i  \bnabla_{\perp}
\phi_G  = \sigma_G$. This allows to re-use common solvers (see Appendix)
for this type of problem to obtain $\phi_G$ and thus $\phi$ from known
$\varepsilon_i$ and $\sigma$. For variable gyroradii $\rho_s ({\bf x})$  or
multiple polarizable species other forms of numerical solvers may have to be implemented. 

The isothermal gyrofluid model consists basically of eqs.~(\ref{eq:continuity})
for both electron and ion gyrocenter densities, which are coupled via
eq.~(\ref{eq:pol-tiff}). The numerical implementation in non-dimensional form
is achieved by the usual drift normalization.
For this purpose the flux divergence contributions in
eqs.~(\ref{eq:continuity}) are first restated.

\subsection{Divergence of perpendicular fluxes}

The divergence of the E-cross-B flux part in eq.~(\ref{eq:continuity})
provides the advective derivative term ${\bs U}_E \cdot \bnabla N_s = (1/B)
[\phi_s,N_s]$ as the primary turbulent nonlinearity, and in the case of an
inhomogeneous magnetic field the ``curvature'' term $N_s \bnabla \cdot {\bs
  U}_E = N_s [\ln B, \phi_s ]$. (Side note: this contribution is here referred to as
``curvature'' although in a strict sense a quasi-2d model with straight
magnetic field lines ${\bs B } = B(x) {\bs e}_z$ only has a gradient-B effect; however, in a toroidal
system both curvature and gradient-B contributions occur in combination,  and can be
treated similarly within joint 3d expressions in (gyro-) fluid models if $T_{\parallel}
\approx T_{\perp}$.)

The 2d advective drift operators $({\bs e}_z \times \bnabla \phi_s) \cdot
\bnabla f \equiv [\phi_s, f]$ are here expressed in Poisson bracket notation,
where $[a,b]=(\partial_x a) (\partial_y b) - (\partial_x b) (\partial_y a)$.
The 2d gyrofluid potential field thus has the meaning of a stream function for the
turbulent E-cross-B flows.

The divergence of the gradient-B flux gives the diamagnetic curvature term
$\bnabla \cdot (N_s{\bs U}_B) = (N_s T_s / q_s B) [\ln B, N_s]$. In contrast
to delta-f models, these terms are not linearized and the full gyrocenter
densities are in this form retained as multipliers to the Poisson brackets.

\subsection{Parallel closure}

In the present quasi-2d model the parallel velocity ${\bs U}_{\parallel}$ contribution can be
approximated by means of the Hasegawa-Wakatani closure \cite{Hasegawa83}.
From the full-f electron parallel momentum equation \cite{Madsen13} in the
quasi-stationary limit a relation in the form of a generalized Ohm's law is
obtained as $e \eta_{\parallel} J_{\parallel} = T_e \nabla_{\parallel} \ln N_e
- e \nabla_{\parallel} \phi$. With $J_{\parallel} \approx - e N_e {\bs
  U}_{\parallel e}$, by assumption of a Spitzer resistivity $\eta_{\parallel} =
0.51 m_e \nu_e /(n_e e^2)$, and applying that the electron gyrocenter density
$N_e \approx n_e$ can be approximated well by the electron particle density
$n_e$, an expression for the term $\bnabla \cdot (N_e {\bs
  U}_{\parallel e}) \equiv - \Lambda_{ce}$ in eq.~(\ref{eq:continuity}) can be obtained
\cite{Held18}. The ion velocity contribution in the parallel response can be
neglected because of the high ion inertia compared to electrons. 

For this purpose a full-f non-adiabatic coupling parameter
$\alpha \equiv T_e k_{\parallel}^2 / (\eta_{\parallel} e^2 n_0 \omega_0)
= n_e T_e k_{\parallel}^2  /(0.51 m_e \nu_e n_0 \omega_0)$ can be defined for
a selected parallel wavenumber $k_{\parallel}$ with $\omega_0=eB/m_i$.
The electron collision frequency $\nu_e$ is in principle proportional to $n_e$,
inversely to $T_e^{3/2}$, and to the (in general density and temperature
dependent) Coulomb logarithm. The usual approximation of a constant Coulomb
logarithm is applied here, and in the present isothermal model only the
density dependence $\nu_e (n_e) \sim n_e \sim N_e$ needs to be discussed \cite{Held18}:
in the classical delta-f fluid HW model the collision frequency is assumed
constant, so that $\alpha$ is a free constant parameter, whereas for the
present full-f model the dependence is kept as $\alpha = N_e \alpha_0$. 
The final non-adiabatic ``ordinary'' HW drive term for electrons in the full-f
form \cite{Held18} is $\Lambda_c = \alpha n_0 \omega_0 [ (e \phi/T_e) - \ln (N/\langle N
\rangle)]$. The angled brackets denote a zonal average, which in the present
2d geometry amounts to averaging in the $y$ direction.

\subsection{Normalization to dimensionless form}

The preceding evaluation of perpendicular and parallel fluxes gives the
(still dimensional) density equations (\ref{eq:continuity}) alternatively as 
\begin{equation}
  \partial_t N_s  + \frac{1}{B} [\phi_s, N_s] + \frac{N_s}{B} [B, \phi_s] +
  \frac{N_s T_s}{ q_s B^2} [B, N_s] = \Lambda_{cs},
  \label{eq:cont2}
\end{equation}
where the nonadiabatic coupling term $\Lambda_{cs}$ only is contributing for 
electrons ($\Lambda_{ci} \equiv 0$).

Time $t$ in the partial time derivative is normalized with respect to $L_{\perp} / c_0$, where $c_0 =
\sqrt{T_e/m_i}$ is the thermal speed, and $L_{\perp}$ is a ``typical'' perpendicular
length scale.  
For local pressure gradient driven systems this is usually set equal to the background
gradient length, here $L_{\perp} \equiv L_n = | \partial_x \ln n_0(x) |^{-1}$, so that temporal
normalization relates to the diamagnetic drift frequency $\omega_{\ast} =  c_0 / L_n$.
For global gradient driven systems often the minor torus radius $L_{\perp}
\equiv a$ is rather used. In case of model systems
with absent background gradient, such as for interchange driven ``blob'' 
setups, $L_{\perp} \equiv \rho_0 \equiv \sqrt{T_e m_i}/(eB)$ is chosen as the
drift scale, and time normalization is related  
to the ion gyration frequency $\omega_0 = c_0 / \rho_0$. These choices can be
set by specifying $\delta \equiv \rho_0 / L_{\perp}$ as a free input parameter.

The perpendicular spatial derivatives in $x$ and $y$ are always normalized with
respect to $\rho_0$, so that $\hat \partial_t \equiv (L_{\perp}/c_0)
\partial_t$, and $\{\cdot,\cdot\} \equiv \rho_0^2 [\cdot,\cdot]$.
Densities and the magnetic field are normalized to reference quantities $\hat
N = N / N_0$ and $\hat B = B / B_0$, respectively, and the electric potential as $\hat 
\phi \equiv e \phi / T_e$. Temperature ratios are defined as $\tau_s \equiv T_s/(T_e Z_s)$.

Dividing eq.~(\ref{eq:cont2}) by $N_s$ and multiplying with $L_{\perp}/c_0$ gives:
\begin{equation}
   \hat \partial_t \ln \hat N_s  + \frac{1}{\hat B \delta}  \{ \hat \phi_s,
   \ln \hat N_s \}   = \hat \Lambda_{cs}   + \hat \Lambda_{Bs} ,
  \label{eq:cont3}
\end{equation}
The normalized dissipative coupling term
$\hat \Lambda_{ce} \equiv (L_{\perp}/c_0) (\Lambda_{ce}/\hat N_e) = \hat
\alpha [ \hat \phi - \ln(N_e/\langle N_e \rangle) ] $,
with $\hat \alpha = \delta (L_{\perp}/L_{\parallel})^2(\omega_{ce}/\nu_e) 
(\hat k_{\parallel}^2 / 0.51)$ as a free parameter, appears only in the
electron equation, and  $\Lambda_{ci} \equiv 0$. The ``modified HW'' model for toroidally more concordant
zonal flow treatment requires to use $\hat \Lambda_{ce} = \hat \alpha [ (\hat
\phi - \langle \hat \phi \rangle ) - ( \ln N_e - \langle \ln N_e \rangle ) ]$
instead \cite{Held18}. In these forms both are directly consistent with their respective common
delta-f limits ($\ln N_e \rightarrow \tilde N_e / N_0$).

In the present quasi-2d setup it is assumed that the 
magnetic field has locally only a weak dependence in $B(x)$, so that $
\delta^{-1} \{ \ln B, f \} \equiv  - \kappa  \hat \partial_y f \equiv - {\cal{K}} (f)$,
with $\kappa \equiv - \delta^{-1}(\hat \partial_x \ln B)$ taken as a constant parameter.
This is also consistent with the assumption of a constant gyroradius,
throughout the computational domain which is supposed to have a radial
extension $L_x\ll R$ much smaller than the major torus radius $R$. For global
simulations across the whole torus cross section this condition would need to
be relaxed.

\begin{figure} 
\includegraphics[width=12cm]{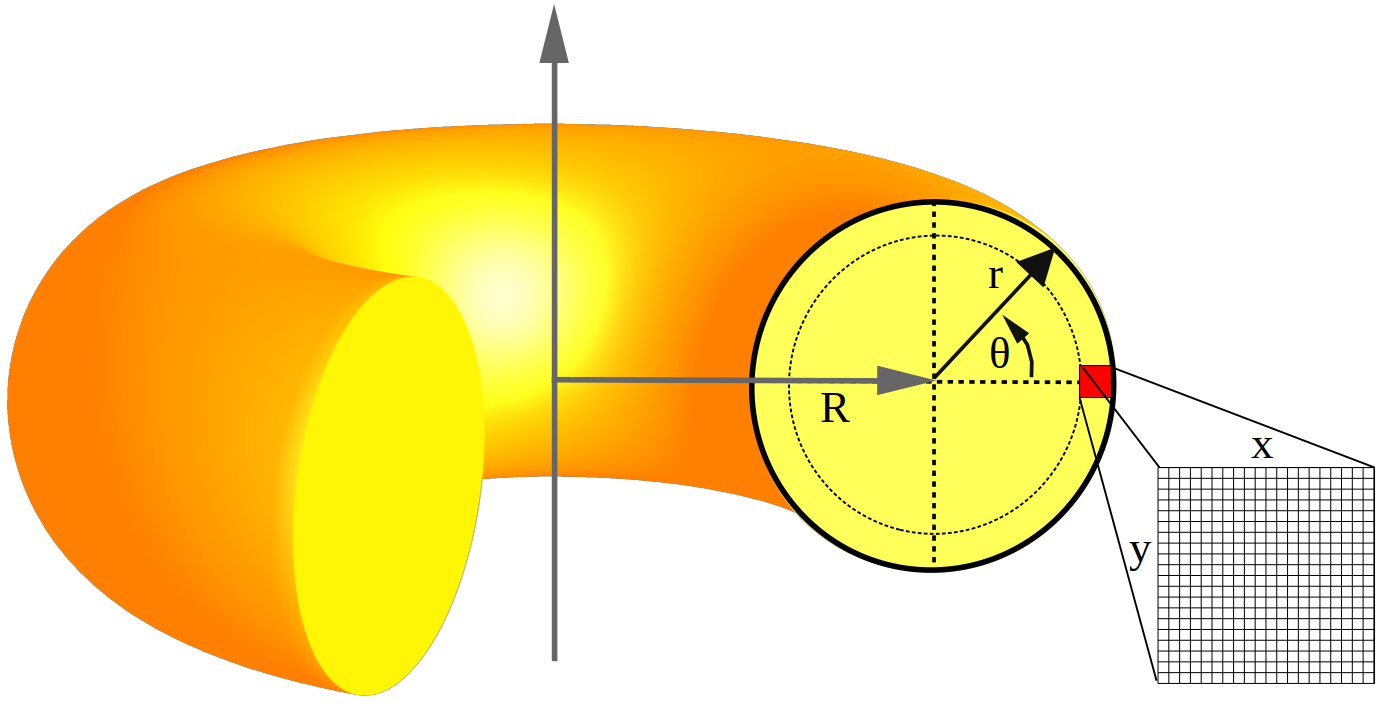} 
\caption{The TIFF simulation domain is set in a small 2d locally
  rectangular ($x$, $y$) area in the poloidal cross-section of a
  torus, such as of a tokamak fusion plasma. The magnetic field direction is
  locally perpendicular to the ($x$, $y$) plane. The advecting fluid-like
  drift velocities in this perpendicular direction are responsible for
  instabilities and plasma turbulence in the presence of (radial) pressure
  gradients and of magnetic field inhomogeneity.}  
\label{f:torus}
\end{figure}

The $x$ direction is here chosen to be local radially outwards on the outboard
midplane side of a torus (compare Fig.~\ref{f:torus}), where $\kappa >0$, and
$\hat B(x) \approx 1 - \kappa x  \approx 1 $ is weakly decreasing with
$x$. This gives  $\hat \Lambda_{Bs} \equiv {\cal{K}} (\hat \phi_s) + \tau_s{\cal{K}} (\ln \hat N_s)$.
The curvature strength can be evaluated as $\kappa \approx 2 \rho_0 / R$.
The temperature ratio is always $\tau_e = -1$ for electrons, and a free
parameter for ions in the order of $\tau_i \sim + 1$. 
The ion temperature ratio $\tau_i$ thus controls the FLR effects, and in
addition here also the ion diamagnetic contribution to gradient-B
(interchange) drive for inhomogeneous magnetic fields ($\kappa \neq 0$).

The formulation in terms of $\ln \hat N$ as the dynamical variable for the
time evolution of densities ensures that $\hat N$ is always positive definite,
which is a requirement for the solution of the generalized Poisson equation,
and directly reduces to the delta-f density equation as $\ln \hat N \approx
\hat N - 1 = (N_0 + \tilde N )/ N_0 - 1 = \tilde N / N_0$ for small relative
fluctuation amplitudes $\hat N \ll N_0$.

\subsection{Normalized polarization}

In a quasi-neutral magnetized plasma under fluid drift ordering, the divergences of
the above perpendicular and parallel fluxes are balanced by the divergence of
the polarization drift. In gyrofluid and gyrokinetic models this is taken into
account by enforcing quasi-neutrality $\sum_s q_s n_s \equiv 0$, and replacing
the particle densities $n_s$ herein in terms of their respective gyrocenter and
polarization densities, with the need to determine a consistent electric
potential as a result.

The polarization equation (\ref{eq:pol-tiff}) is made non-dimensional by
applying the same normalizations as above, which results in 
$  \hat \bnabla \cdot \sqrt{\Gamma_0} \hat \varepsilon_i \sqrt{\Gamma_0}  \hat
  \bnabla_{\perp} \phi  = \hat \sigma$,
with  $\hat \varepsilon_i =  \hat N_i / \hat B^2$
and  $\hat \sigma = -  \sum_s  Z_s \Gamma_{1s} \hat N_s ({\bf x})$.
The gyration operators are already dimensionless by definition.

\subsection{Selection of edge and scrape-off-layer model scenarios}

The usual applications for quasi-2d simulations of nonlinear drift dynamics in
magnetized plasmas are, for example, fundamental studies on either (a) gyrofluid turbulence
and zonal flows with FLR effects, or on (b) FLR effects on
interchange dynamics of warm ``blob'' perturbation propagation. Both cases are
commonly treated in separate studies, where for (a) $\Lambda_{Bs} \equiv 0$,
and instead for (b) $\Lambda_c \equiv 0$ is set respectively, for better
separability and understanding of basic underlying mechanisms.

``Blob'' transport is regarded as most relevant in the scrape-off-layer (SOL)
of fusion plasmas. Studies with single or few seeded blobs are important to
reveal fundamental mechanisms, but in tokamaks blobs are presumed to be generated rather
``randomly'' around the separatrix, so that turbulent drift wave vortices and 
zonal flow effects from the closed field line (CFL) edge region will play a
large role for consistent studies of ``blobby'' (intermittent) SOL
transport. Several 2d fluid codes (for example as in
refs.~\cite{Halpern16,Dudson17,Madsen16}) take into account both drift wave
and interchange effects and different background conditions by assigning two
regions, that are defined by a ``separatrix'' location $x_s$ within the same 2d
computational domain.   

An additional effect that should be taken into account in the SOL region is
sheath coupling of the open field lines with limiter or divertor material
walls. This effect intrinsically includes parallel kinetics along the magnetic
field direction. Gyrokinetic, gyrofluid and fluid models for sheath coupling
conditions may be approximated under assumption of Bohm conditions. Any
perpendicular 2d fluid-type approximation will likely miss relevant physics here, but
can in principle be included also in the present 2d full-f gyrofluid
model. For test purposes presently a (partly inconsistent) sheath instability
``toy model'' is optionally included in the TIFF code with additional coupling
terms added to the right hand side of eq.~(\ref{eq:cont3}), which needs
further improvement before any application. 
For example, a 2d delta-f form of the gyrofluid sheath coupling terms was
introduced by Ribeiro \cite{Ribeiro08}, as 
$ N_e \hat \Lambda_{Se} \rightarrow \gamma_D [(1+\Lambda_D) \tilde N_e- \hat \phi]$
and $\hat \Lambda_{Si} \rightarrow \gamma_D \tilde N_e$, with sheath coupling
parameter $\gamma_D$ as defined in ref.~\cite{Ribeiro08}, and $\Lambda_D =
\log(\sqrt{m_i/2 \pi m_e})$. This model add-on in principle will allow to
examine simplified edge-SOL coupled turbulence and flow dynamics including
turbulent generation of SOL blobs, once a consistent full-f full-k sheath
coupling term is derived and implemented in the future. For this reason, the
present paper only discusses simulations without sheath coupling.

All coupling terms $\hat \Lambda$ on the right hand side of eq.~(\ref{eq:cont3})
are then selectively only applied in the respective regions of interest, for
example formally by multiplication with Heaviside type step functions
$\lambda(x_s)$ for a given relative separatrix position $0 \leq (x_s/L_x) \leq 1$.

Application cases and systematic physics studies will be presented
elsewhere. Here the focus is on introduction of the code TIFF and its
presently underlying model as a reference for later applications and
further developments, such as toroidal geometry and inclusion of thermal and
electromagnetic dynamics in a (field-aligned) 3d extension.
A main aspect here is also on introduction and testing of an efficient dynamically
corrected Fourier solver for the generalized Poisson problem in the polarization.

\section{Numerical solution algorithm}
\label{sec:numerical}

The normalized equations solved in the TIFF code are:
\begin{eqnarray}
  \hat \partial_t \ln \hat N_e + \{ \hat \phi, \ln \hat N_e \} 
  &= & \lambda_B \hat \Lambda_{Be} + \lambda_S  \hat \Lambda_{Se} + \lambda_c
       \hat \Lambda_c\label{eq:tiff-dense}  \\  
  \hat \partial_t \ln \hat N_i + \{ \hat \phi_i, \ln \hat N_i \} 
  &= &   \lambda_B \hat \Lambda_{Bi}  + \lambda_S \hat \Lambda_{S i}  \label{eq:tiff-densi} \\ 
  \hat \bnabla \cdot  \hat \varepsilon_i  \hat \bnabla_{\perp} \hat \phi_G   &=& \hat \sigma_G
                                                                 \label{eq:tiff-pol}
\end{eqnarray}

The general procedure for solution of this set of equations is as follows:

\noindent (\#1) Specify $\hat N_e({\bs x})$ and $\hat N_i({\bs x})$ on an equidistant grid in a 2d
rectangular ($x,y$) domain, either as initial condition, or subsequently
updated in each time step by eqs.~(\ref{eq:tiff-dense}-\ref{eq:tiff-densi}). 

\noindent (\#2) Compute $\hat N_{Gi}  \equiv \Gamma_{1i} \hat N_i $. The
constant gyroradius assumption allows to evaluate all gyro-averaging
operations efficiently in ${\bs k}$ space, here $\hat N_{Gi}({\bs k}) = \hat N_i({\bs
  k})/(1+\tau_i \hat k^2/2)$. 
In the TIFF code presently the 2d discrete Fourier transform from the FFTW3
library \cite{fftw3} is used for transformations between ${\bs k}$ and ${\bs
  x}$ space representations. 

\noindent (\#3) Apply boundary conditions (see next section) on $\hat N_e$, $\hat N_i$ and $\hat N_{Gi}$.

\noindent (\#4) Prepare the input functions to eq.~(\ref{eq:tiff-pol}) as $\hat \sigma_G
= \sqrt{\Gamma_0}^{-1} \hat \sigma = \sqrt{1+\tau_i\hat k^2} \; \hat \sigma$ with
$\hat \sigma = \hat N_e - Z_i \hat N_{Gi}$, and $\hat \varepsilon_i = \hat N_i/\hat B^2$.

\noindent (\#5)  Obtain $\hat \phi_G$ from eq.~(\ref{eq:tiff-pol}) with one
of the solvers discussed in the Appendix.

\noindent (\#6) Compute the electric potential from  $\hat \phi = \sqrt{\Gamma_0}^{-1}
\hat \phi_G$ also via ${\bs k}$ space.

\noindent (\#7) Compute the gyrofluid ion potential $\hat \phi_i = \Gamma_{1i} \hat
\phi +  |\hat \nabla_{\perp} \hat \phi_G |^2 / (2 \hat B^2)$.

\noindent (\#9) Update $\hat N_e$ and $\hat N_i$ in time through
eqs.~(\ref{eq:tiff-dense}-\ref{eq:tiff-densi}) and return to step (\#1). 

The time step update (\#9) first requires evaluation of the advective Poisson brackets and
all coupling terms $\hat \Lambda$.
The brackets $\{ \cdot, \cdot \}$ are here presently solved with the energy
and enstrophy conserving (but not shock capturing) fourth order Arakawa scheme
\cite{Arakawa66,Naulin03}. 

In $\hat \Lambda_{Bs}$ the curvature operators ${\cal K} (f) = \kappa \hat
\partial_y f$ are evaluated by (fourth order) centered finite differencing
over the (periodic) $y$ direction.
Evaluation of $\hat \Lambda_{Ss}$ and $\hat \Lambda_{ce}$ are
straightforward. $\hat \Lambda_{ce}$ includes calculation of zonal averages
$\langle f \rangle (x) = (1/n_y) \sum_j f_{i,j}$, which on a 2d rectangular
local grid simply requires summation over all $n_y$ grid points $j$ of the $y$ direction.

Time step updating of $f \equiv \ln \hat N_s$ in the form $\partial_t f = F$ here uses
a fourth order accurate three-step Adams-Bashforth method with Karniadakis
weights \cite{Karniadakis91}: 
\begin{eqnarray}
  f^{(t+1)} &= &  c_0 f^{(t)} - c_1 f^{(t-1)} + c_2 f^{(t-2)} \nonumber \\ 
  &+& c_F \Delta t \left[ 3 F^{(t)} - 3 F^{(t-1)} +
F^{(t-2)} + \hat \Lambda_{\nu}^{(t)} \right]
\label{eq:karniadakis}
\end{eqnarray}
where $c_0 = 18/11$, $c_1 = 9/11$, $c_2 = 2/11$ and $c_F = 6/11$.
The (normalized) time step size $\Delta t$ has to be small enough for CFL stability.
For further numerical stabilization of the otherwise explicit scheme an
artificial sub-grid type viscosity term $\Lambda_{\nu}$ is added.
For this a hyperviscosity $\hat \Lambda_{\nu} = - \nu_4 \hat \bnabla^4 \ln
\hat N_s$ is here applied. The coefficient $0 \leq \nu_4 \ll 1$ is
heuristically chosen to prevent grid instability at smallest scales, as a sink
for the nonlinear direct vorticity cascade, and to (slightly) damp out Gibbs type
noise that can appear at under-resolved strong gradients due to the not shock
capturing nature of the Arakawa discretization. The value of $\nu_4(h)$ needs to
be adapted when spatial grid resolution $h$ is changed to achieve optimum
results. If desired, for example for comparability to other code implementations, an
ordinary ``physical'' viscous diffusion term $\hat \Lambda_{\mu} = + \mu (\hat  \bnabla^2
\hat N_s) / \hat N_s$ could be added to the right hand side of
eqs.~(\ref{eq:tiff-dense}-\ref{eq:tiff-densi}). For usual application of the
gyrofluid model to hot fusion edge plasmas the actual viscosity $\mu$ in general
would be too small to be resolved efficiently by direct numerical simulation.

For solution of the generalized Poisson problem as in step (\#5) of the
outlined algorithm, presently three methods are implemented in TIFF as
described in more detail in the Appendix: 
An iterative preconditioned conjugate gradient (PCG) solver, an iterative red-black
successive over-relaxation (SOR) solver, and a novel dynamically corrected
Fourier (DCF) solver. All solvers make use of the results for $\phi^{(t-1)}$
and  $\phi^{(t-2)}$ of the previous time steps, either for extrapolated 
initialisation of the iterative schemes, or for the (therefore denoted
``dynamical'') correction of the approximate Fourier method. 
The DCF method leads to predictable run times of the generalized Poisson
problem, that only depend on the grid size but not on other simulation
parameters, whereas the number of iterations and run time can vary
significantly for the PCG and SOR solvers for any specified accuracy, in
particular when densities (and thus $\hat \varepsilon_i({\bf x})$) are strongly inhomogeneous.

\section{Initial and boundary conditions}

Initially the gyrocenter density $\hat N_i(x,y)$ and $\hat N_e(x,y)$ fields
have to be specified. For full-f initial background profiles a radially exponential
decline is set, each with $\hat N_{0}(x) = \hat N_L \exp(-x / d)$ for $d = L_p / \ln(n_L/N_p)$.
Here $L_p = x_s$ denotes either the width of the pedestal region if a
separatrix at $x_s<L_x$ is applied, or the width $L_x$ of the whole $x$ domain
if no SOL region is treated. The initial $\ln \hat N_s$, profiles (and in its limit
the delta-f $\tilde N$ profiles) are thus linear.
On top of this initial background, either single perturbations, such as a
Gaussian density blob, or a pseudo-random ``bath'' of modes, is added with a
given amplitude.
The initial ion gyrocenter density is either set equal to the electron
density, or a ``vorticity free'' initialisation with
$\hat N_i = \Gamma_1^{-1} \hat N_e$ is used so that $\hat \sigma =0$.

Boundary conditions are applied in several instances: on density profiles in order to maintain
a background gradient if required, for the intrisic boundary value problem of
the Poisson equation, for solution of the gyro operators in ${\bs k}$ space, and on finite
difference operators in ${\bs x}$ space.
The ``poloidal'' $y$ direction is assumed periodic. The box length $L_y$ in units of
$\rho_0$ has to be chosen much larger than typical perpendicular correlation lengths for
turbulence simulations, or much larger than blob or vortex scales for simulation
of such structures, in order to minimize self-interaction effects across the
periodic $y$ boundary.

\subsection{Density profile boundary conditions}

Density profiles could be maintained by specifying source terms  $\hat
\Lambda_Q(x)$ around the inner ($x=0$) radial boundary and sink terms around
the outer ($x=L_x$) boundary.

The presently considered application scenarios are for example a full-f
gyrofluid generalization of HW turbulence simulations, or seeded blob
simulations on a (usually) constant background density. For comparability to
common delta-f implementations of these scenarios it is adequate to maintain an
average density profile by prescribing fixed boundary densities $\hat N(x=0)
\equiv \hat N_L$ and $\hat N(x=L_x) \equiv \hat N_R$, while the full densities
may still self-consistently evolve in between. Seeded blob simulations
could for example use $\hat N_L = \hat N_R = 1$.

``Classical'' fluid or delta-f gyrofluid HW turbulence simulations usually decouple a constant
background gradient in the advective derivative as $ \delta^{-1} \{
\phi,\tilde N +N_0(x) \} = \delta^{-1} \{ \phi,\tilde N \} + g \hat \partial_y \phi$ with
a gradient parameter $g = \delta^{-1} (\rho_s/L_n)$. For the ``diamagnetic
drift frequency'' normalization of time this amounts to $g=1$, but $g$ could
also be kept as a free parameter \cite{Numata07}. 
The restriction in delta-f fluid or gyrofluid simulations to small
fluctuations that are decoupled from the background profile enables the use of
periodic boundary conditions in $x$ on $\tilde N$ and $\phi$. This is not feasible
any more for both delta-f or full-f simulations with global profile evolution.

If $\delta$ is specified as an input parameter, then this needs to be chosen
consistently with the required density boundary values for profile driven
simulations as $\delta = (\hat N_L - \hat N_R)/L_x$. When for example the
radial domain size is $L_x = 64$ in units of $\rho_0$, then setting $\hat N_L =
1 + 0.5$ and $\hat N_R = 1 - 0.5$ specifies $\delta =1/64 \approx 0.015$.

The typical experimental tokamak edge steep density gradient pedestal regions,
which are the usual scenario of interest for HW model simulations, have widths
in the order of around 50 to 100 $\rho_0$ and drift scales $\delta \sim {\cal O} (10^{-2})$.

\subsection{Parametric transition to delta-f limit within full-f equations}
\label{transition}

Specification of the density boundary in this way allows for a consistent treatment
and verification of the delta-f limit of small perturbation amplitudes on
large relative background densities within the full-f code. For gradient
driven turbulence this can be achieved by reducing all of the background
variation, drift scale, and initial perturbation amplitudes by the same small
factor $\epsilon$. For example, setting $\hat N_L = 1 + \epsilon \cdot 0.5$
and $\hat N_R = 1 - \epsilon \cdot 0.5$ for the same box size of $L_x = 64$
gives $\delta \approx \epsilon \cdot 0.015$. When initial (for example blob) perturbation
amplitudes $\epsilon \cdot \Delta n$ are in the full-f model chosen smaller by the same
factor (for example $\epsilon = 1/100$) as in a corresponding delta-f code
(with initial amplitude $\Delta n$), then this corresponds to the respective
delta-f setup and enables a direct comparison (and code cross-verfication) of
this limit when $\epsilon \rightarrow 0$. 

\subsection{Radial boundary density condition}

It is desirable to reduce density fluctuations and maintain zero vorticity
and/or flows on narrow inner and outer radial boundary layers. The boundary
layer region of ``width'' $L_{\beta} \ll L_x$ is here defined by a function $\beta(\hat x) =
1 - a_L \exp[-\hat x^2/L_{\beta}^2] - a_R \exp[-(1-\hat x)^2/L_{\beta}^2]$
with $\hat x = x / L_x$. The parameters $a_L$ and $a_R$ are set to $1$ for
boundary gradient driven cases (as described above), or can be respectively
set to $0$ if source and/or sink terms $\hat \Lambda_Q(x)$, or a free outflow
condition on the outer boundary, are activated.

The vorticity around the radial boundaries can be approximately set to zero,
when equivalently the right hand side $\hat \sigma = \hat N_e({\bf x}) -  Z_i
\Gamma_{1i} \hat N_i ({\bf   x})$ of the polarization equation, corresponding
to the first order polarization density with FLR effects, is set to zero.

For this purpose it is most convenient to first define the boundary values of the gyroaveraged ion
gyrocenter density  $\hat N_{Gi}$ and then compute the consistent electron and
ion gyrocenter densities in each time step ($\#3$ in the algorithm), by re-setting
$\hat N_{Gi}(x,y)  = \Gamma_{1i} \hat N_i (x,y)\rightarrow [ \hat N_{Gi}(x,y)-
\hat N_{0}(x) ] \cdot \beta(\hat x) + \hat N_{0}(x)$. This leaves the bulk
density unchanged and gives a smooth radial transition to the initial
profile values $\hat N_{0}(x)$ only in narrow regions of width $L_{\beta}$.
The corrected ion gyrocenter density is then computed by $\hat N_i =
\Gamma_{1i}^{-1} \hat N_{Gi}$ in ${\bf k}$ space and re-set in the boundary
region.  The electron density is re-set to $\hat N_{e} \rightarrow [ \hat
N_{e} - \hat N_{Gi} ] \cdot \beta(\hat x) + \hat N_{e}$, which ensures zero
$\hat \sigma$ in the boundary region. A further correction could be
applied to alternatively ensure zero E-cross-B vorticity $\hat \Omega =
\hat \bnabla_{\perp}^2 \hat \phi$ at the boundaries, but tests have shown no
significant changes or advantages, as the E-cross-B vorticity is reduced
already jointly with the generalized vorticity at the boundaries in this
approach. (In the delta-f limit both cases are identical.)

This condition $\hat \sigma|_{b.c.} \equiv 0$ avoids strong vorticity
gradients at the $x$ boundaries (and by this  possibly related numerical
instabilities), and in addition ensures well-defined solution of the
polarization equation (see Appendix). 

Existence of a unique solution to the generalized Poisson problem
$\bnabla \cdot {\bf P} = \bnabla \cdot \varepsilon \bnabla \phi = \sigma$
requires fulfillment of a compatibility boundary condition. Taking the domain
integral over the $(x,y)$ area $S$, one has $\int_S d{\bf x} \;  \bnabla \cdot
{\bf P} = \int_S d{\bf x} \; \sigma$, and therefore $\int_{\delta S} dl \; {\bf P}
\cdot {\bf   n}_S= \int_S d{\bf x} \; \sigma$. This is ensured by the above
vorticity-free and flow-free (${\bf n}_S \cdot \bnabla \phi = \partial_x \phi
= u_y \equiv 0$) conditions on the boundary $\delta S$ in $x$ (and periodicity in $y$).
The condition correponds to global conservation of the polarization charge density.

\subsection{Mirror padding for Fourier transforms}

The algorithm involves four evaluations of gyro-operators (in steps
$\#2,3,4,6$), which is in the present isothermal model achieved with high
accuracy in ${\bf k}$ space by applying (FFTW3 library) Fourier transforms.
Actually the evaluation of $\hat N_i = \Gamma_{1i}^{-1} \hat N_{Gi} = [ 1 -
(1/2) \tau_i \bnabla^2] \hat N_{Gi}$ in Pad\'e approximation for the vorticity
free boundary condition (step $\#3$) could alternatively be achieved by
(considerably faster but less accurate) finite differencing in ${\bf x}$
space. Although more costly, this evaluation is here for consistency also done in ${\bf k}$ space.

Standard Fourier transformation requires periodic boundary conditions, else
discontinuities would introduce Gibbs noise artefacts.
The radial ``physical'' domain $0 \leq \hat x \leq 1$ however usually includes
density profiles with $\hat N_s (\hat x = 0) \neq \hat N_s(\hat x = 1)$.
For this reason, ghost domains in an extended $\hat x$ direction are introduced for
the $(x,y)$ field arrays before solution by Fourier transforms to ${\bf k}$ space,
and also before solution of the polarization equation.

Doubling the (initially ``quarter-wave'' physical) domain to $0 \leq \hat x
\leq 2$, by copying the initial array $f(\hat x, \hat y)$ into the region $1 < \hat x \leq 2$ and
defining symmetrically mirrored $f (\hat x) \equiv \hat f(1 - \hat x)$
for $0 \leq \hat x \leq 1$ ensures (``half wave'') $\hat x$ periodicity of field functions.
Full-wave input functions to the gyro-averaging Fourier transforms are achieved by a
further anti-symmetric domain doubling with boundary offset correction: the
four-fold extended array includes $f(4 - \hat x) = 2 f(0) - f (\hat x)$ for $2
< \hat x \leq 4$. This ensures full-wave radial representation of the
densities in the gyro-averaging Fourier transforms, but for a four-fold
computational cost.

It is as usual favourable to use power-of-two numbers of grid points in the
$\hat x$ and $\hat y$ directions, so that the FFTW3 library will make use of
Fast Fourier Transform (FFT) algorithms. Any other grid size can be chosen, but
FFTW3 then automatically uses somewhat (depending on the grid size) slower
Discrete Fourier Transform (DFT) methods. 

\section{Parallelization and reproducibility}

The present development and production run platforms for the TIFF code are
multi-core shared memory office workstations, so that multi-core and/or multi-thread
parallelization is simply achieved by OpenMP (OMP) parallelization of grid array
sized double loops, and use of the respective OMP library of FFTW3 for the
Fourier transforms. 
The frequent transistions between single and parallel regions, required by the
above algorithm, prevent good scaling. Efficient speed-up is usually achieved
(depending on the input parameters and on the hardware) when between 8 and 32
parallel threads are used.

Bitwise reproducibility of subsequent executions on the same system can be
achieved when the ``ESTIMATE'' flag for the transform planner routine of the
FFTW3 library is used, which is desirable for verification and testing
purposes. The ``MEASURE'' flag is faster in execution (which would be
desirable for long production runs with for example millions of time steps)
but not regularly bitwise reproducible \cite{fftw3}, because it might choose
different algorithms depending on the system background load, which can be quickly noticeable in the
turbulent phase of simulations due to the highly nonlinear nature of the
present set of equations. This may not be relevant if anyhow only statistical
diagnostic quantities of a saturated fully turbulent state are of interested.

The TIFF code presently uses only the FFTW and OMP libraries and can be compiled
for example with the GNU gcc compiler, which are all available as standard in
most usual Linux distributions. This ensures usability on most consumer PCs or
workstations.

\section{Diagnostic output}
\label{sec:diagnose}

Diagnostic outputs are produced only every $n_D$-th time steps. For $n_D$
large (in the order of hundreds or thousands of time steps) this 
allows real-time diagnostics with little post-processing, which greatly
reduces the necessary output storage space (in comparison to writing all arrays
for each computational time step). 

The 2d dynamical field arrays $\hat N_e(i,j)$, $\hat N_i(i,j)$, $\hat \phi(i,j)$, $\hat
\omega(i,j)$, $\hat \sigma(i,j)$, and $\tilde N_e(i,j) = \hat N_e(i,j) - \hat
N_0(i)$ are regularly written out (e.g. to a RAM disc), and can be viewed
(already during run time) with any 2d visualization software (for which
presently a simple gnuplot script is used). In addition, 1d cross sections of
several arrays $f(i,j_0)$ and $f(i_0,j)$, and averaged radial profiles
$\langle f \rangle_j (i)$ are written. 
 
Spectra $\langle f(k_x) \rangle_j$ and $\langle f(k_y) \rangle_i$ are computed
at each diagnostic output step for several quantities, such as kinetic energy,
enstrophy, and density and potential power spectra.

Energetic and transport quantities are recorded as time traces of ($x,y$) domain averages.
The (normalized) E-cross-B advective electron particle transport is obtained as
$Q_n(t) = (1/S) \int d {\bf x} \; \hat N_e(x,y) \partial_y \hat \phi(x,y)$, where $S= L_x Ly$,
and the $y$ derivative is provided by simple centered finite differencing
(as also in the following diagnostics).

The full-f global thermal free energy is given by (compare ref.~\cite{Held23}):
$E_T = E_{Te} + \tau_i E_{Ti}$ with 
$E_{Ts} = (1/S) \int d {\bf x} \; [ \hat N_s ( \ln \hat N_s - \ln \hat N_0 ) - (\hat N_s -  N_0)]$.
The kinetic energy is $E_K = (1/2S) \int d {\bf x} \; \hat N_i (\hat \bnabla \hat \phi_G)^2$.
The total energy is an ideal conserved quantity, and
can be used as a diagnostic for saturation of a turbulent state.

Further diagnostics can be included, for example output and computation of
difference norms for solver testing against a constructed solution, or
center-of-mass calculation of an interchange driven ``blob''.

\section{Delta-f limit}
\label{deltaf}

The TIFF code also implements the corresponding 2d isothermal delta-f set of
gyrofluid equations concurrent to the full-f model. Where possible both use
the same procedures, or enter forks for specific treatment. This allows
cross-verification of the full-f model in its small-amplitude limit with the
original delta-f model within the same code and for equal methods and (initial
and boundary) conditions.

The equivalently normalized delta-f set of equations is:
\begin{eqnarray}
  \hat \partial_t \hat N_e + \{ \hat \phi, \hat N_e \} \;
  = & \lambda_c \hat \Lambda_c  \;  +&  \lambda_B \hat \Lambda_{Be} + \lambda_S
                                       \hat \Lambda_{Se} \label{eq:df-dense}  \\ 
  \hat \partial_t \hat N_i + \{ \hat \phi_i, \hat N_i \} \;
  = &  &  \lambda_B \hat \Lambda_{Bi}  + \lambda_S \hat \Lambda_{S i}  \label{eq:df-densi} \\ 
  \frac{1}{\tau_i}  \left( \Gamma_{0i} -1 \right) \hat \phi    = & \hat \sigma &   \label{eq:df-pol} 
\end{eqnarray}

The (modified) delta-f HW term is $\hat \Lambda_{ce} = \hat \alpha [ (\hat
\phi - \langle \hat \phi \rangle ) - ( \hat N_e - \langle \hat N_e \rangle ) ]$.
Sheath coupling is described by $ \hat \Lambda_{Se} = \gamma_D [(1+\Lambda_D) \tilde N_e- \hat \phi]$
and $\hat \Lambda_{Si} = \gamma_D \tilde N_e$.
The curvature terms are $\hat \Lambda_{Bs} = {\cal{K}} (\hat \phi_s) + \tau_s{\cal{K}} (\hat N_s)$. 
Hyperviscosity equivalently now acts on $\hat N_s$.

In $\hat \sigma =  \hat N_e ({\bf x}) -  Z_i \Gamma_{1i} \hat N_i ({\bf
  x})$ the gyrocenter densities $\hat N_s = \tilde N_s / N_0$ here denote the
fluctuating components only. Equivalently to the full-f version, 
the radial density profile is also included into the initialization of the
densities and evolved accordingly, and is not decoupled by a gradient
parameter, so that the same (not periodic) boundary conditions on the $x$
domain apply as for the full-f case described above. This leads to additional
boundary damping in contrast to doubly periodic (physical) domains, so that
the results can only be qualitatively compared with usual fluid HW code results.

The solution algorithm is basically the same as for the full-f model described
above. Steps $\#$ 4-6 are reduced to evaluating the delta-f (full-k) polarization
eq.~(\ref{eq:df-pol}) in $k$ space with the Pad\'e approximation of $ \Gamma_{0i}$ through $\hat
\phi_k = - [ \hat k^2 /(1 + \tau_i \hat k^2)] \; \hat \sigma_k $. The
long-wavelength (low-k) form of this delta-f polarization is optionally obtained by
setting the term $\tau_i \hat k^2 \equiv 0$ herein. The gyrofluid ion
potential (step $\# 7$) is simply $\hat \phi_i = \Gamma_{1i} \hat \phi$.

For diagnostics, the delta-f total thermal free energy is evaluated as
$E_T = (1/S) \int d {\bf x} [ ( \hat N_e -\hat N_0 )^2 + \tau_i (\hat N_i -
\hat N_0)^2]$ and kinetic energy as $E_K = (1/2S) \int d {\bf x} \; (\hat
\bnabla \hat \phi_i)^2$, while the transport and other output quantities
remain the same as for the full-f model. 

\section{Test of  generalized Poisson solvers}

In step $\#$ 5 of the full-f TIFF algorithm, the polarization equation in the
form of a generalized 2d Poisson problem $\bnabla \cdot  \varepsilon  \bnabla \phi  = \sigma$
has to be solved for the unknown $\phi$. Equivalent problems arise in other
physical scenarios, such as for spatially variable dielectric with
permittivity $\varepsilon ({\bf x})$ and a given (negative) space charge
distribution $\sigma ({\bf x})$. Iterative 
solution algorithms are routinely applied on such problems, but in most
applications the inhomogeneity in $\varepsilon$ is usually small, whereas it
can vary strongly in our case of a turbulent edge plasma. Iterative
solvers may converge slowly in such situations, if at all.

The present TIFF code implementation includes a choice between an iterative preconditioned
conjugate gradient (PCG) solver, an iterative red-black successive
over-relaxation (SOR) solver, and a novel ``dynamically corrected'' Fourier
(DCF) solver. The solution algorithms are described in detail in the
appendix. Other solvers are of course possible, perhaps faster, and may be further implemented
and tested in future. The full-f gyrofluid FELTOR code for example also includes a
discontinuous Galerkin method or a multi-grid scheme, in addition to a
(different) conjugate gradient scheme \cite{FeltorV6}. 

Here the PCG, SOR and DCF solvers in TIFF are compared by method of a constructed
solution as a component test, and within the full code setup by
cross-verification with recent ``blob'' dynamics results from the FELTOR code.
The main aspect here lies on (cross) verification and determination of
accuracy of the new DCF scheme, which is suggested as a stable and
(predictably) efficient solution method for application on the
generalized Poisson problem of the full-f polarization equation, that is
applied in each (small) time step of a dynamically evolving turbulence code. 

The dynamical context is important, because the underlying ``Teague method'',
introduced in different context, for approximate solution of a generalized
Poisson-type problem, is in itself not very accurate. Here however an efficient dynamical
(or recursive) correction based on the solution from the previous time step is newly applied
with the ``Teague method'', which is shown to improve its accuray by two
orders of magnitude, and by that can achieve similar accuracy as iterative
solvers with a ``reasonable'' (affordable) number of iterations in a turbulence code.

\subsection{Teague's original method}

In optics, the ``transport of intensity equation'' (TIE) is an approximate relation for the
intensity and the wave phase of a coherent beam in an optical field
\cite{teague83,zuo20}. The underlying mathematical form of the TIE is
basically equivalent to the 2d generalized Poisson problem: $\bnabla \cdot I({\bf
  x}) \bnabla \psi({\bf x})  = -k \partial_z I({\bf x})$, where $I$ is the
known (measured) planar intensity distribution at a distance $z$, $k$ the wave number, and
$\psi$ the sought phase of the wave. Various solution methods have been
used on the TIE in the field of applied optics \cite{zuo20}, but a now widely used
approximate Fourier method has been suggested by Teague in 1983
\cite{teague83} by introducing what is now commonly known as Teague's auxiliary function. 

In the following this original ``Teague's method'' is described in terms of
the notation introduced in the previous sections in context of the gyrofluid
polarization equation (and not in the TIE notations used in optics). Starting
with
\begin{equation}
  \bnabla \cdot  [ \varepsilon ({\bf x}) \bnabla \phi ({\bf x}) ] = \sigma  ({\bf x}),
  \label{eq:otm-gpol} 
\end{equation}
an auxiliary 2d scalar function $p({\bf x})$ is introduced that is supposed to satisfy
\begin{equation}
  \varepsilon ({\bf x}) \bnabla \phi ({\bf x}) \equiv \bnabla p ({\bf x}).
  \label{eq:otm-aux} 
\end{equation}
This reduces the generalized Poisson problem to an ordinary 2d Poisson equation
\begin{equation}
  \bnabla^2 p ({\bf x}) = \sigma ({\bf x})
\end{equation}
which can, for example, be efficiently inverted and solved by Fourier transformation into ${\bf k}$ space
(and back-transformation of the solution) as $p({\bf k}) = - (1/k^2) \sigma
({\bf k})$, or by any other fast Poisson solver.
The defining auxiliary relation eq.~(\ref{eq:otm-aux}) can be re-written as
$  \bnabla \phi ({\bf x}) = [ \bnabla p ({\bf  x}) ] / \varepsilon ({\bf x})$,
on which the divergence operator is applied on both sides to obtain:
\begin{equation}
  \bnabla^2 \phi ({\bf x}) = \bnabla \cdot \frac{1}{\varepsilon ({\bf x}) } 
  \bnabla p ({\bf x}).
  \label{eq:otm-phi} 
\end{equation}
The right hand side contains the (by now) known quantities $\varepsilon ({\bf
  x})$ and $p ({\bf x})$, and can be evaluated by standard fourth order 2d
centered finite differencing in $x$ and $y$. Eq.~(\ref{eq:otm-phi}) can
therefore again be solved (e.g. in ${\bf 
  k}$ space) to obtain $\phi ({\bf x})$. When Fourier transforms are used, the
method involves four evaluations (two forward and two backward) of the 2d
transform ${\cal F}$, which can be formally expressed (compare
ref.~\cite{paganin98} for the TIE version) as:
\begin{equation}
  \phi ({\bf x}) = {\cal F}^{-1} \left [ - \frac{1}{ k^2} {\cal F} \; \bnabla \cdot \frac{1}{\varepsilon} 
  \bnabla {\cal F}^{-1} \left ( - \frac{1}{ k^2}  {\cal F} \sigma   \right) \right].
\end{equation}

Apparently the accuracy of Teague's method is considered mostly sufficient for solution
of the TIE in optics, as it is widely applied. However, it has of course been noted that
the introduction of the auxiliary function $p$ in eq.~(\ref{eq:otm-aux}) is in
general mathematically incomplete, and this approximation may introduce an
unspecified error into the solution \cite{schmalz11,ferrari14}. The definition of eq.~(\ref{eq:otm-aux})
would actually hold exactly for a truly conservative nature of the vector
field ${\bf P}  ({\bf x}) \equiv \varepsilon ({\bf x}) \bnabla \phi ({\bf x})$, but this
condition of irrotationalty is in general not ensured, neither in the TIE of
optics, nor in polarization of electrostatics.
 
Rather, the complete Helmholtz decomposition of the vector field  ${\bf P}({\bf x})$ is:
\begin{equation}
 {\bf P}  =  \varepsilon \bnabla \phi \equiv  \bnabla p + \bnabla \times {\bf H}
  \label{eq:otm-helm} 
\end{equation}
with a scalar potential $p({\bf x})$ and a vector potential ${\bf H}({\bf x})$.
This general form provides the same solution path for $p({\bf x})$ like above, as still $\bnabla \cdot
{\bf P} = \bnabla^2 p = \sigma$ holds. The next step, generalizing the result
of eq.~(\ref{eq:otm-phi}), now gives:
\begin{eqnarray}
  \bnabla^2 \phi ({\bf x}) &=& \bnabla \cdot \frac{1}{\varepsilon ({\bf x}) } 
  \bnabla p ({\bf x}) \; + \; \bnabla \cdot \frac{1}{\varepsilon ({\bf x}) }
                               \bnabla \times {\bf H} ({\bf x}) \nonumber \\
  &=& \bnabla \cdot \frac{1}{\varepsilon ({\bf x}) } 
  \bnabla p ({\bf x}) \; + \; \left \{ \frac{1}{\varepsilon}, \eta \right \}
  \label{eq:otm-phi-gen} 
\end{eqnarray}
with $\bnabla \cdot (1/\varepsilon) \bnabla \times {\bf H} = (\bnabla \times
{\bf H}) \cdot \bnabla (1/\varepsilon) =\{ (1/\varepsilon), \eta \}$, where the 2d
Poisson bracket notation (defined above) is used, and $\eta$ is the $z$
component of (unknown) ${\bf H}(x,y) = \eta (x,y) \; {\bf e}_z$.

To find a constraint on $\eta$, one can apply the curl operator on both sides,
instead of the divergence operator $\bnabla \cdot \bnabla \phi = \bnabla
\cdot ...$, correspondingly as in eq.~(\ref{eq:otm-phi-gen}). This gives the
condition $\bnabla \times \bnabla \phi = \bnabla \times  (1/\varepsilon)
\bnabla p + \bnabla \times [ (1/\varepsilon) \bnabla \times {\bf H}] = 0$,
resulting in:
\begin{equation}
  \bnabla \cdot  \left( \frac{1}{\varepsilon} \bnabla \eta \right) = \left \{
    \frac{1}{\varepsilon}, p \right \}.
  \label{eq:otm-eta}
\end{equation}
This determining relation for the unknown $\eta$ however again has the form of a
generalized Poisson equation, which sends us back to start...

Another constraint can be generated by taking the curl on ${\bf P}$, which
gives the relation $\bnabla \times ( \varepsilon \bnabla \phi ) = \bnabla \times
\bnabla \times {\bf H}$. This can be rephrased as:
\begin{equation}
  \bnabla^2 \eta  = \{ \phi, \varepsilon \}.
  \label{eq:otm-rec}
\end{equation}
This first shows that a sufficient condition for the accuracy of Teague's
method would be that $\{ \phi, \varepsilon \} = \bnabla \phi \times \bnabla
\varepsilon \equiv 0$, which holds if the isocontours of $\phi$ 
and $\varepsilon$ were aligned everywhere \cite{schmalz11}. The relation can
also be used to derive constraints for the error norm \cite{schmalz11}.

If applied to gyrofluid simulations, the error is surely not negligible as
$\{ \phi, \varepsilon \} \sim \{ \phi, N_i \}$ for constant magnetic field, and thus
directly related to the advective ion nonlinearity.

\subsection{Iterative and dynamical corrections}

Iterative methods have been suggested to correct the error by the original
Teague approximation. In ref.~\cite{zuo14} a Picard-type iteration is
applied, that uses an initial solution $\phi_0$ by Teague's method
(without need to refer to the vector potential) to compute an according source
term $\sigma_0 \equiv \bnabla \cdot \varepsilon \bnabla \phi_0$, then use
$\Delta \sigma = \sigma_0 - \sigma$ in another turn of application of Teague's
method to obtain a correction $\Delta \phi$, and repeat until $\Delta \phi$ is
smaller than a specified error.  

The set of equations (\ref{eq:otm-phi-gen}) and (\ref{eq:otm-rec}) allows
a further possibilities of an iterative approach. A first approximation for
$\phi^{(0)}$ is obtained by setting $\eta^{(0)} = 0$ in eq.~(\ref{eq:otm-phi-gen}),
which corresponds to the original Teage approximation; then compute an
approximate $\eta^{(1)}$ from exact solution of eq.~(\ref{eq:otm-rec}) using
this approximate $\phi^{(0)}$, and with that obtain an updated  $\phi^{(1)}$
from solution of eq.~(\ref{eq:otm-phi-gen}); and iterate further until a desired error
bound is reached.

In the following, this recursive correction (here denoted as ``RCF method'')
based on eqs.~(\ref{eq:otm-phi-gen}) and (\ref{eq:otm-rec}) will be tested by
means of a constructed solution. However, any such iterative methods involve
multiple evaluations of Fourier transforms (or other fast conventional Poisson
solvers) per iteration step, and can become expensive when the iterative
evaluation has to be carried out in every of very many time steps in a
dynamical turbulence simulation. 

It will be shown that already after one or two iterations a high accuracy is
reached. This motivates another possible non-iterative correction to Teague's
method in the context of a dynamical simulation with time evolution of
$\phi(t)$ in the presence of small time steps.  
In simulations of fully developed turbulence, and in particular when an
explicit finite difference scheme is used like in the present code, the time
step $\Delta t$, as it appears in eq.~(\ref{eq:karniadakis}), is small, and so
accordingly is the difference between successive solutions $\phi^{(t)}$ and $\phi^{(t-1)}$. 

The idea is to therefore evaluate
\begin{equation}
  \eta_o  ({\bf x}) \equiv  \bnabla^{-2} \{ \phi^{old}, \varepsilon \}
\end{equation}
with an ``old'' solution from previous time steps in a dynamical simulation,
and with this approximation compute 
\begin{equation}
  \phi ({\bf x})  = \bnabla^{-2} \left[ \bnabla \cdot \frac{1}{\varepsilon } 
  \bnabla \left (  \bnabla^{-2} \sigma \; \right ) + \; \left \{ \frac{1}{\varepsilon}, \eta_o
  \right \} \right].
\end{equation}
In contrast to iterative schemes applied within each time step, this reduces
the additional expense for the correction to only one (for example FFT based)
inversion of the conventional Poisson problem, and (comparatively cheap)
computation of another Poisson bracket. 
In the very first time step one can simply use $\eta_o = 0$ and thus obtain an
uncorrected approximate solution for $\phi$ by the conventional Teague method.
It is feasible to simply use $\phi^{old} \equiv \phi^{(t-1)}$ of the previous time
step. In the present code an extrapolation is applied as:
\begin{equation}
\phi^{old} \equiv \phi^{(t-1)} + a \cdot (\phi^{(t-1)} - \phi^{(t-2)} )
\end{equation}
with a free estimator factor $a \in (0, 1)$. Because this extension of
Teague's method uses previous results of a time evolving simulation, and
evaluates the multiple occuring inversions of the Poisson problem with Fourier
solvers in ${\bf k}$ space, it is in the following abbreviated as the ``DCF''
(dynamically corrected Fourier) method. The accuracy however will here depend on
the size of the (small) dynamical time step.

\subsection{Unit test of generalized Poisson solvers}

The specific implementation of the PCG, SOR and DCF solvers as used in the
TIFF code is in detail described in the Appendix. The code can be run with a
unit testing option, by setting a flag in the input parameter file, that
initialises analytical constructed functions $\varepsilon_c$ and $\sigma_c$ and
only calls the specified Poisson solver once, so that the numerical solutions
for $\phi$ can be directly compared with the analytical function  $\phi_c$.
The main purpose of this test is to determine the applicability and accuracy of the novel dynamically
corrected Fourier (DCF) solver in comparison to the established SOR and PCG methods.

As constructed solutions, here $\phi_c \equiv \phi_{c0} \sin ( x k_x )  \sin ( y k_y )$ and
$\varepsilon_c \equiv 1 - g x + a \sin (x k_n) \sin (y k_n)$ are exemplarily specified,
and the corresponding $\sigma_c = \bnabla \cdot \varepsilon_c \bnabla \phi_c =
\varepsilon_c (\partial_x^2 \phi_c + \partial_y^2 \phi_c ) + (\partial_x
\varepsilon_c )(\partial_x \phi_c ) + (\partial_y \varepsilon_c )(\partial_y \phi_c )$ 
is also given analytically. In the following test, $\phi_{c0} = 1$, $a=0.2$, $g
= 0$ or $1$, $k_x = 2 (2 \pi/L_X)$, $k_y = 3 (2 \pi / L_y)$ and $k_n = 4 (2
\pi /L_x)$ are set. 

\begin{figure} 
\includegraphics[width=8cm]{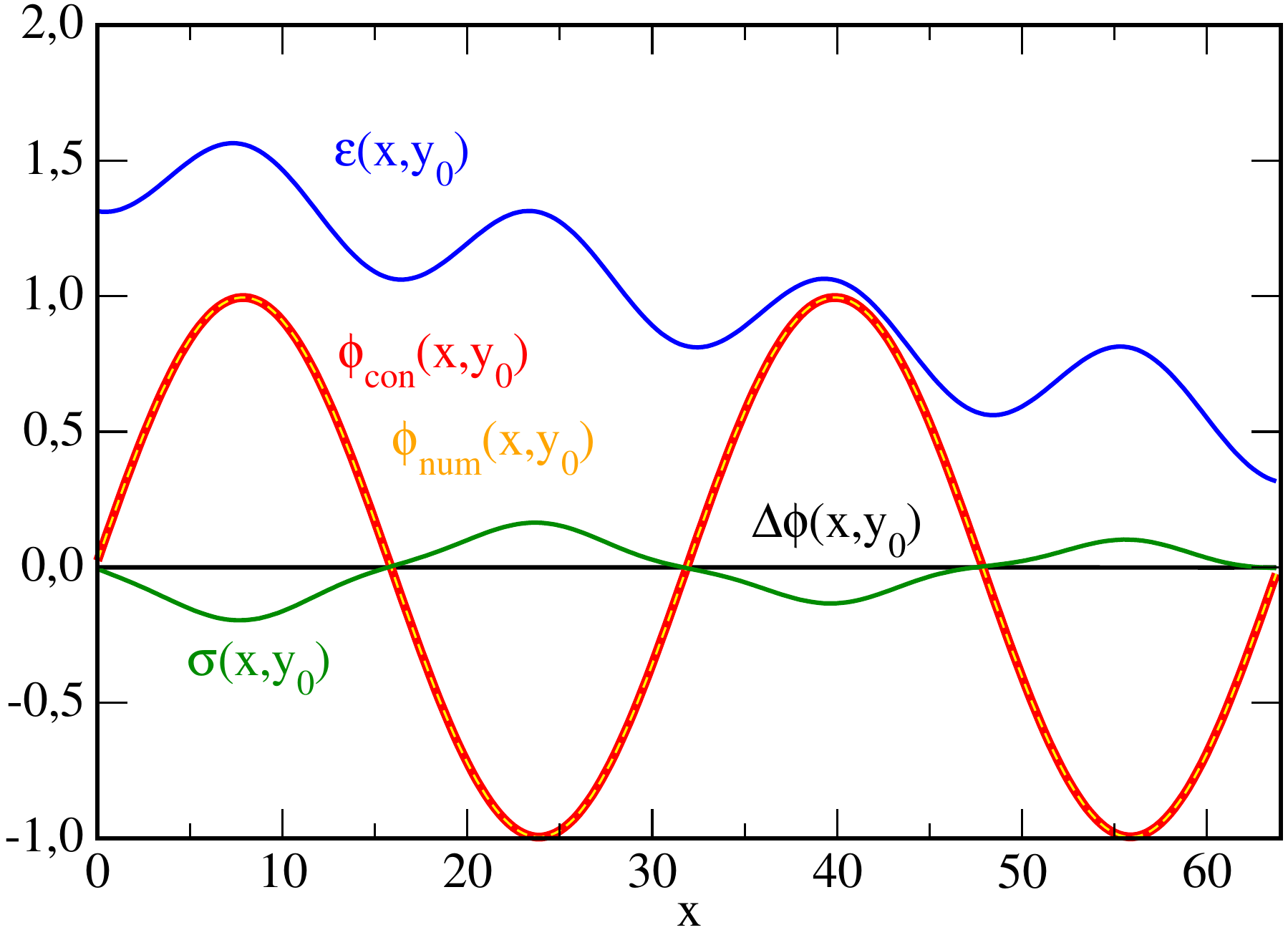}
\includegraphics[width=8cm]{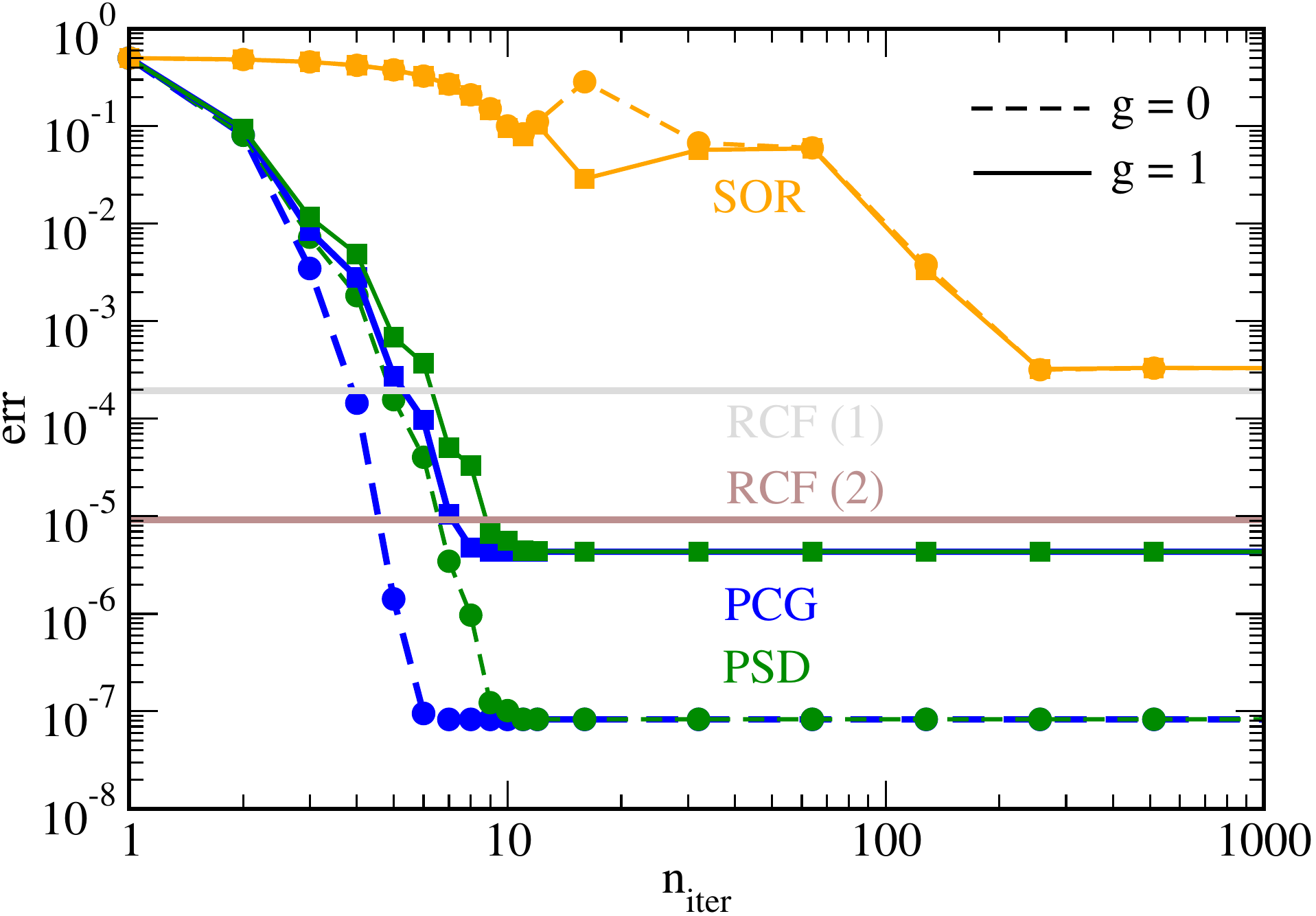}
\caption{Left: Constructed solution $\phi_{c}$ (red) and $\varepsilon_{c}$
  (blue) and corresponding $\sigma_c = \bnabla \cdot  \varepsilon_c \bnabla
  \phi_c$ (green), shown as a function of $x$ at $y_0 = (L_y/2 -5)$. The
  difference $\Delta \phi$ (black) from the numerical solution (dashed 
  orange) obtained with the RCF(1) method for $N_x = N_y = 256$ is here too small
  to be visible.
Right: Global $L_2$ error norm obtained from different solvers (see main text)
after $n_{iter}$ iterations with ($g=1$, solid) and without ($g=0$, dashed) a background gradient
in  $\varepsilon_{c}$. 
} 
\label{f:error_iter}
\end{figure}

The constructed and numerical solutions are shown in Fig.~\ref{f:error_iter}
(left) as a function of $x$ at $y_0 = (L_y/2 -5)$. The differences $\Delta
\phi(x,y)$ between constructed and numerical solutions can be visualised and
compared by their global $L_2$ norms as a function of resolution or iterations. 

In Fig.~\ref{f:error_iter} (right) the global error  $err = ||\Delta \phi||_2$
is generally large for $g=1$ when a background gradient in $\varepsilon_c
(x)$ (solid lines) is present and discontinuities in the derivatives occur at
the $x$ boundaries. For comparison, also results for the above constructed 
solution but with $g=0$ (constant background, periodic $\varepsilon_c (x)$,
dashed lines) is shown, with reduced more ``ideal''  errors. In practice, the
$g=0$ case would for example correspond to a ``seeded blob'' simulation
scenario, whereas the $g=1$ case would be relevant to gradient driven drift wave turbulence scenarios.

The error $err$ is shown as a function of the number of iterations $n_{iter}$
for different solvers, on a square equidistant grid with $N = N_x = N_y =
256$. For reference the two horizontal bold lines denote the error for $g=1$ of the
recursively corrected Fourier (RCF) method after one recursion (grey line) and
after two recursions (brown line). The dynamically corrected Fourier solver
(DCF) uses recursion by means of the value $\phi_0$ from the last dynamical
time step and therefore can not be quantified with a unit test, but one might
expect an error of the DCF solver in between the RCF(1) and RCF(2)
solvers. For higher recursion numbers ($\geq 4$) the error of the RCF solver 
approaches the error of the high $n_{iter}$ limits of the PCG (blue, solid
line) and PSD (green, solid line) solvers, as all of those schemes employ
Fourier methods and fourth order finite differencing. (The pre-conditioned
steepest descent (PSD) solver is described in the Appendix in context of the
PCG solver.)
The error (for $g=1$) of the PCG scheme is after $n_{iter} = 5$ iterations
comparable to the RCF(1) scheme. The computing times of RCF and PCG are here for
this PCG iteration number also similar.

The following table depicts the $L_2$ error and order for the $g=0$ test case
above, given with the RCF(4) scheme with four recursive iterations as a
function of grid resolution $h \sim N_0/N$. The order is calculated as $ O =
\log( ||\Delta \phi||_2^{(h)} / ||\Delta \phi||_2^{(h/2)} ) / \log( 2/1 )$. A
higher number of iterations does not give further improvement in this case. 
The table shows the expected scaling with fourth order accuray, which
corresponds to the order of the finite difference schemes used in the evaluation of the
right hand side terms in eq.~(\ref{eq:otm-phi-gen}).


\setlength{\tabcolsep}{.5cm}
\begin{table}[h]
\begin{tabular}{|c|c|c|}
  \hline
  N    & err & O \\
  \hline
  16   & $1.808 \cdot 10^{-2}$  & - \\
  32   & $1.251 \cdot 10^{-3}$  & 3.85 \\
  64   & $7.998 \cdot 10^{-5}$  & 3.96 \\
  128  & $5.036 \cdot 10^{-6}$  & 3.99 \\
  256  & $3.154 \cdot 10^{-7}$  & 4.00 \\
  512  & $1.972 \cdot 10^{-8}$  & 4.00 \\
  1024 & $1.226 \cdot 10^{-9}$  & 4.01 \\
  2048 & $7.113 \cdot 10^{-11}$ & 4.11 \\
  \hline
\end{tabular}
  \caption{$L_2$ norm error and order of accuracy for various square grid
    point numbers $N= N_x=N_y$ for the constructed solution test case (compare
    Fig.~\ref{f:error_iter}), obtained with the ``recursively corrected
    Fourier'' (RCF) method after 4 iterations.} 
\end{table}


The second order accurate SOR scheme (orange lines in Fig.~\ref{f:error_iter})
has much slower convergence and does for 
this resolution never reach the accuracy of the (fourth order) RCF and PCG
schemes. Usually several hundred SOR iterations are required for acceptable
results in the unit test, where the initial  $\phi (n_{iter}=0) \equiv 0$ for the
iteration is zero. The convergence (for a given error tolerance) 
however improves  drastically within a dynamical simulation, when the initial
$\phi(n_{iter}=0)$ of the iteration is projected from the result $\phi(t-1)$ of the
previous time step.

\begin{figure} 
\includegraphics[width=8cm]{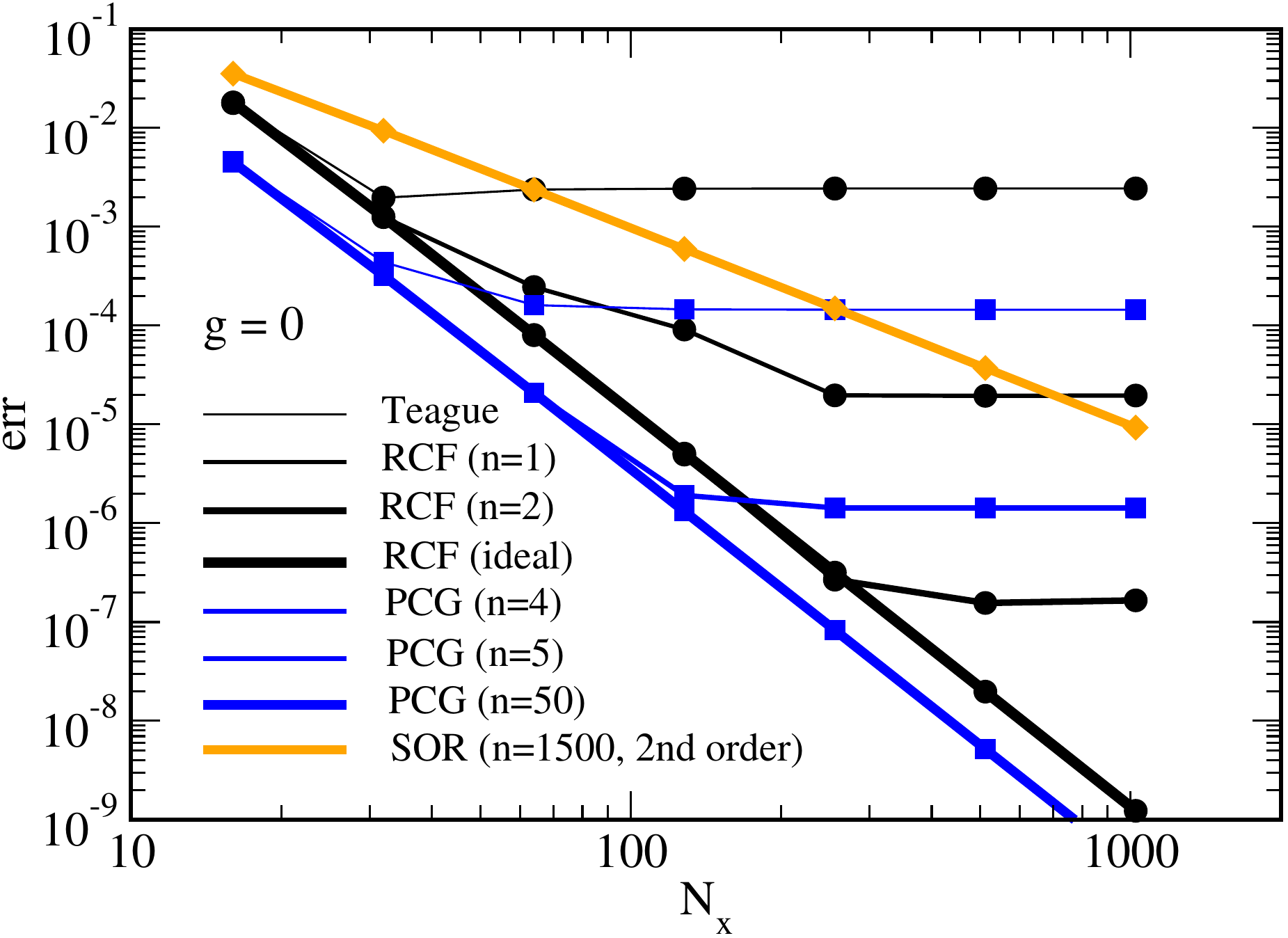}
\includegraphics[width=8cm]{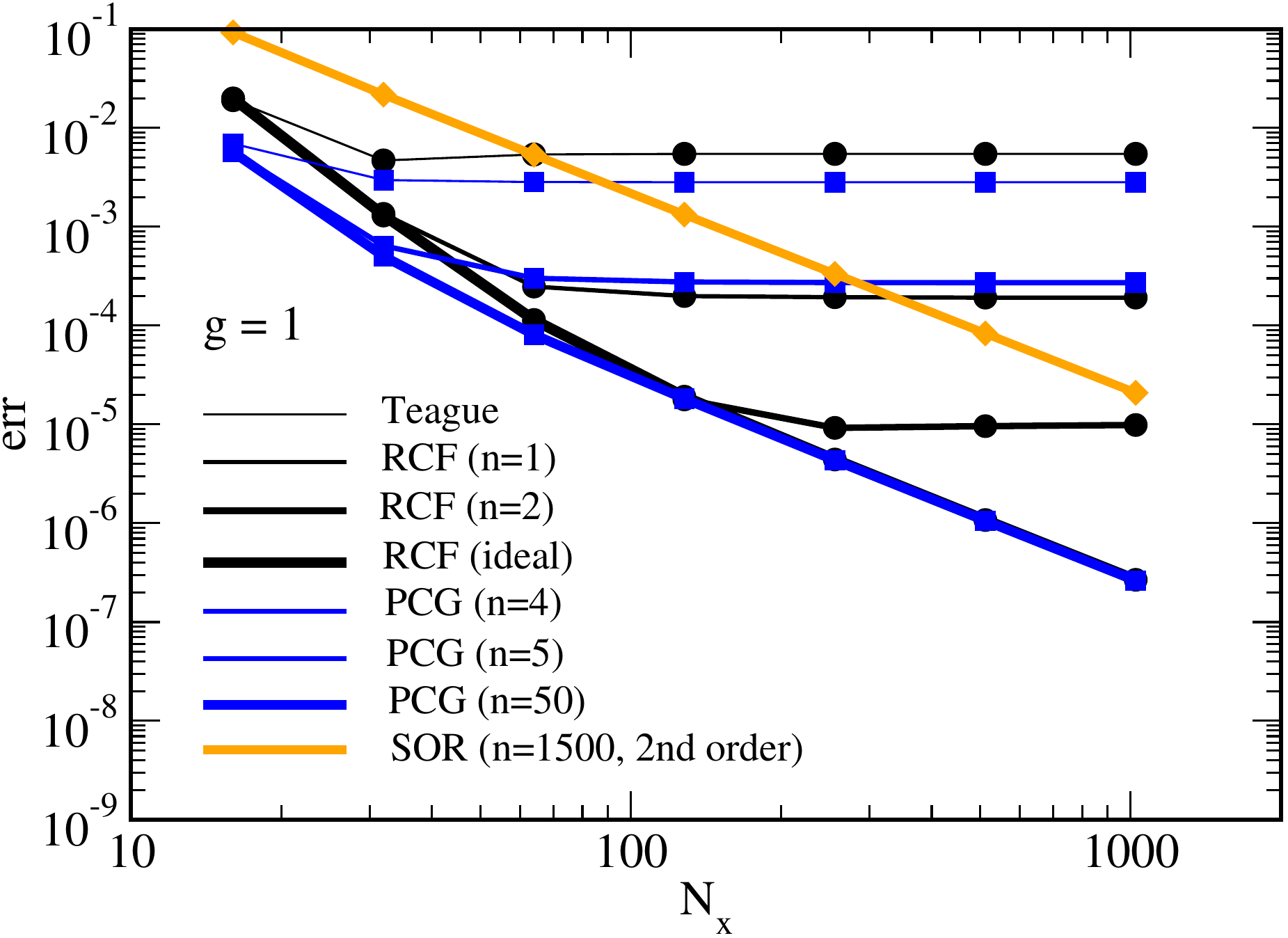}
\caption{$L_2$ error norm, related to the constructed solution as a function of number of
  grid points $N_x = N_y$, for various generalized Poisson solvers (see main
  text). As expected, for $g=0$ (left figure, periodic functions) the
  RCF and PCG solvers show a fourth order dependence (for large iteration
  numbers $n$), and the SOR solver a second order
  dependence. For $g=1$ (right figure, nonperiodic  $\varepsilon_{c}$ with
  background gradient) the errors are larger for all solvers due to boundary
  discontinuity in the $x$ derivatives.} 
\label{f:error_nx_g0}
\end{figure}

In Fig.~\ref{f:error_nx_g0} the error is shown as a function of the resolution
in terms of the number of grid points $N_x = N_y$, with otherwise same
parameters as above. The case of periodic density ($g=0$) is shown in the left
figure: for a large number of iterations ($n_{iter}=50$) the PCG scheme (bold blue
line) follows the expected fourth order dependence on resolution, whereas the
SOR scheme (bold orange line, $n_{iter}=1500$) shows the second order
dependence $err \sim N_x^{-2}$ (i.e., a slope of -2 in the log-log plot). The
``ideal'' RCF scheme (bold black line) here denotes a correction
term calculated once from  the constructed solution  $\phi_c$ instead of
recursively, which also shows a fourth order ($N_x^{-4}$) slope, as expected.

The gradually thicker black lines (from top to bottom) show, topmost, the error of the
original ``Teague method'' without correction, which is for usual resolutions
always in the range of $2 \cdot 10^{-3}$ for $g=0$ (left figure). The RCF(1)
scheme uses one recursion of  the solver in the unit test problem, and already
gives an improvement of the error by around two orders of magnitude. The
RCF(2) scheme with two recursions follows the ``ideal'' dependence until
around $N_x = 256$ and for higher resolution saturates at an error of around
$2\cdot 10^{-7}$.  
The gradually thicker blue lines (from top to bottom) show the error by the PCG scheme after $4$,
$5$ and $50$ iterations. The $n_{iter}=5$ case for PCG has for high resolution an
error in between the results from the RCF(1) and RCF(2) schemes.

The right Fig.~\ref{f:error_nx_g0} shows the corresponding results for $g=1$
in the presence of a background gradient, which in the present implementation
of the solvers in TIFF overall reduces the achievable accuracy.
The general behaviour of the different solvers is similar to the $g=0$
results, but no ``ideal'' scaling of the error (-4 power law) with resolution is obtained any more.
The RCF(1) and PCG(4) schemes also show similar errors for medium to large resolution.

The dynamically corrected Fourier (DCF) scheme needs three evaluations of the
standard Poisson problem, which is here presently achieved by a fast Fourier
solver. In a dynamical simulation the error correction then is calculated from
the previous time step solution. The PCG scheme needs one evaluation of a
standard Poisson inversion (e.g. by Fourier solver) per iteration step, so
that PCG ($n=3-4$) has approximately the same computational cost as a DCF
evaluation (which in addition needs another call of a Poisson bracket evaluation).
The necessary number of iterations in PCG or SOR to reach a given accuracy can 
strongly depend on the complexity of the problem, basically determined by the
degree of variability of $\varepsilon (x,y,t)$. The computational expense of 
the DCF method only depends on the size of the grid. The DCF scheme 
therefore presents itself as a viable alternative method (for moderate accuray) with
predictable run times.

\subsection{Cross-verification of full-f full-k blob simulation}

The unit testing of the PCG, DCF and SOR solvers above has been applied on the
``pure'' generalized Poisson equation $\bnabla \cdot \varepsilon \bnabla \phi
= \sigma$. The implementation in the TIFF code for solution of the full-f
full-k gyrofluid polarization eq.~(\ref{eq:pol-tiff}) requires for $\tau_i \neq
0$ two additional applications of (here Fourier based) solvers for the
$\sqrt{\Gamma_0}$ operator, and further (also Fourier based) evaluations of
the gyro-averaging operator ${\Gamma_1} N_i$ in $\sigma$, and of ${\Gamma_1}
\phi$ in the ion gyrocenter density advection.

As a further test, an overall cross-verification of the code is intended by
running a seeded ``blob'' simulation for parameters that correspond to a
recent study with the full-f full-k gyrofluid isothermal version of the
FELTOR code, described by Held and Wiesenberger in ref.~\cite{Held23} and
shown in Figures 4-7 therein.  

The simulation plasma input parameters for this case are: ion to electron
temperature ratio $\tau_i = 4$, magnetic curvature $\kappa = 1.5 \cdot 10^{-4}$,
normalization scale $\delta = 1$, non-adiabaticity $\hat \alpha = 0$, and
absence of a sheath ($\hat \Lambda_{Ss} = 0$). The square domain is $L_x = L_y =
200 \rho_s$ with resolution $N_X = N_y = 1024$. 
The initial perturbation is a Gaussian ``blob'' in electron density of width
$w = 5 \rho_s$ and peak amplitude $\Delta \hat N = 1$ centered at $x_0 = 25$
and $y_0 = 100$, on a constant background with $\hat N_0 = 1$.
The initial perturbed ion gyrocenter density is either set
equal to the electron density (which introduces an initial blob spin), or for
a ``vorticity free'' initialization to $\hat N_i = \Gamma^{-1} N_e$.
The $x$ boundary density values are pinned in a small zone
$L_{\beta} = 2 \rho_s$. 
The hyperviscosity is set to $\nu_4 = 10^{-5}$, and in accordance to
ref.~\cite{Held23} an additional physical viscosity $\nu_2 = 3 \cdot 10^{-5}$
is applied.
The time step is set to $\Delta \hat t = 0.1$ and run for $I_{max} = 23040$
steps with diagnostic outputs at every $I_{out} = 320$ steps. 
The iterative error bound for the PCG and SOR solvers was set to $err =
10^{-3}$ with 500 iterations maximum.

\begin{figure} 
\includegraphics[width=16cm]{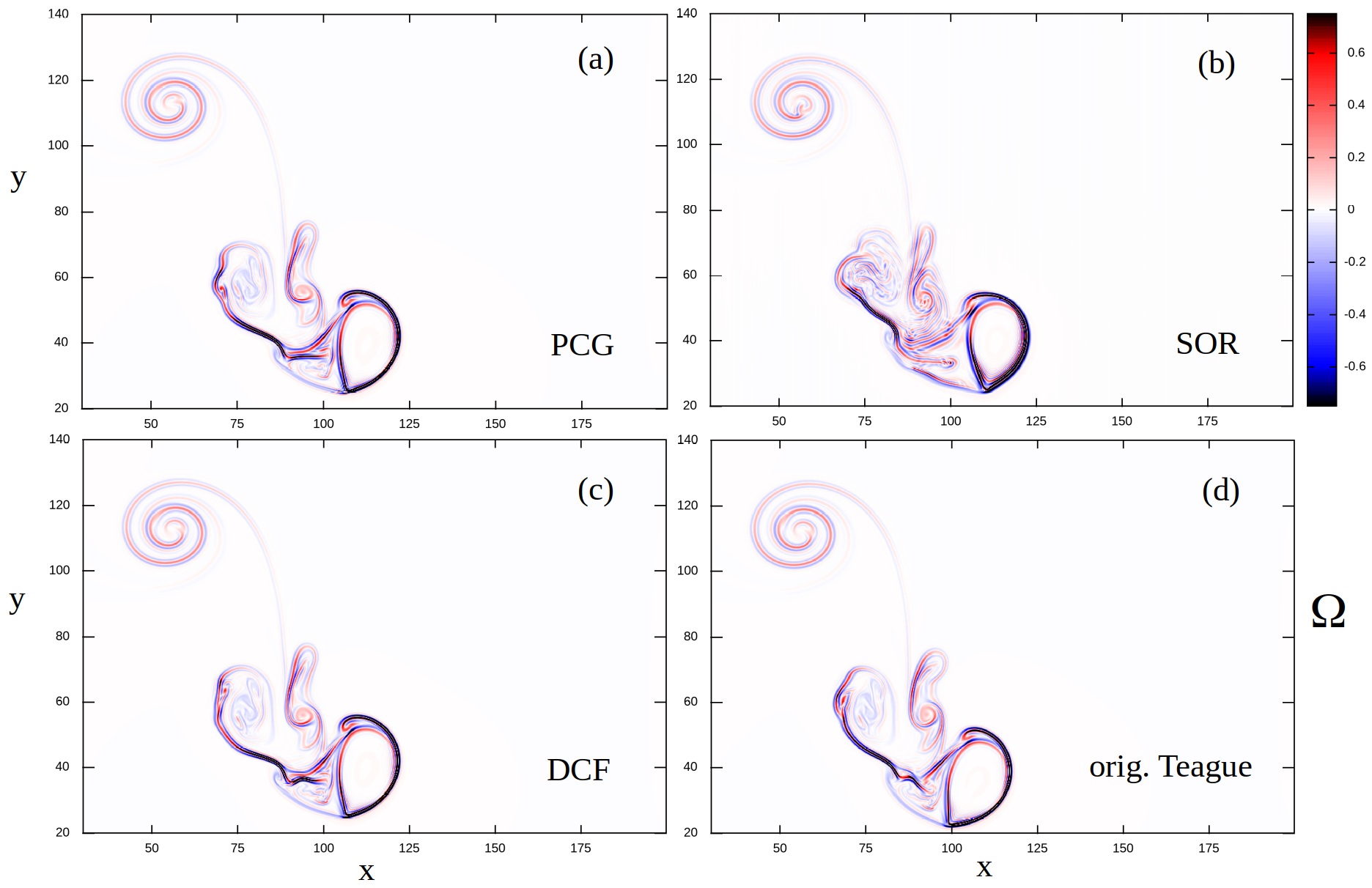} 
\caption{Comparison of vorticity $\hat \Omega = \hat \bnabla^2 \hat \phi(x,y)$
  for an evolved seeded blob ($\tau_i=4$, $\hat N_0=1$, $w=5$) with  zero
  vorticity initial condition ($\hat N_i = \Gamma^{-1} N_e$), at $\hat t=2304$, for different generalized
  Poisson solvers. (a) PCG; (b) SOR; (c) DCF; (d) original (uncorrected) Teague solver.} 
\label{f:blob-compare-vorfree}
\end{figure}

\begin{figure} 
\includegraphics[width=16cm]{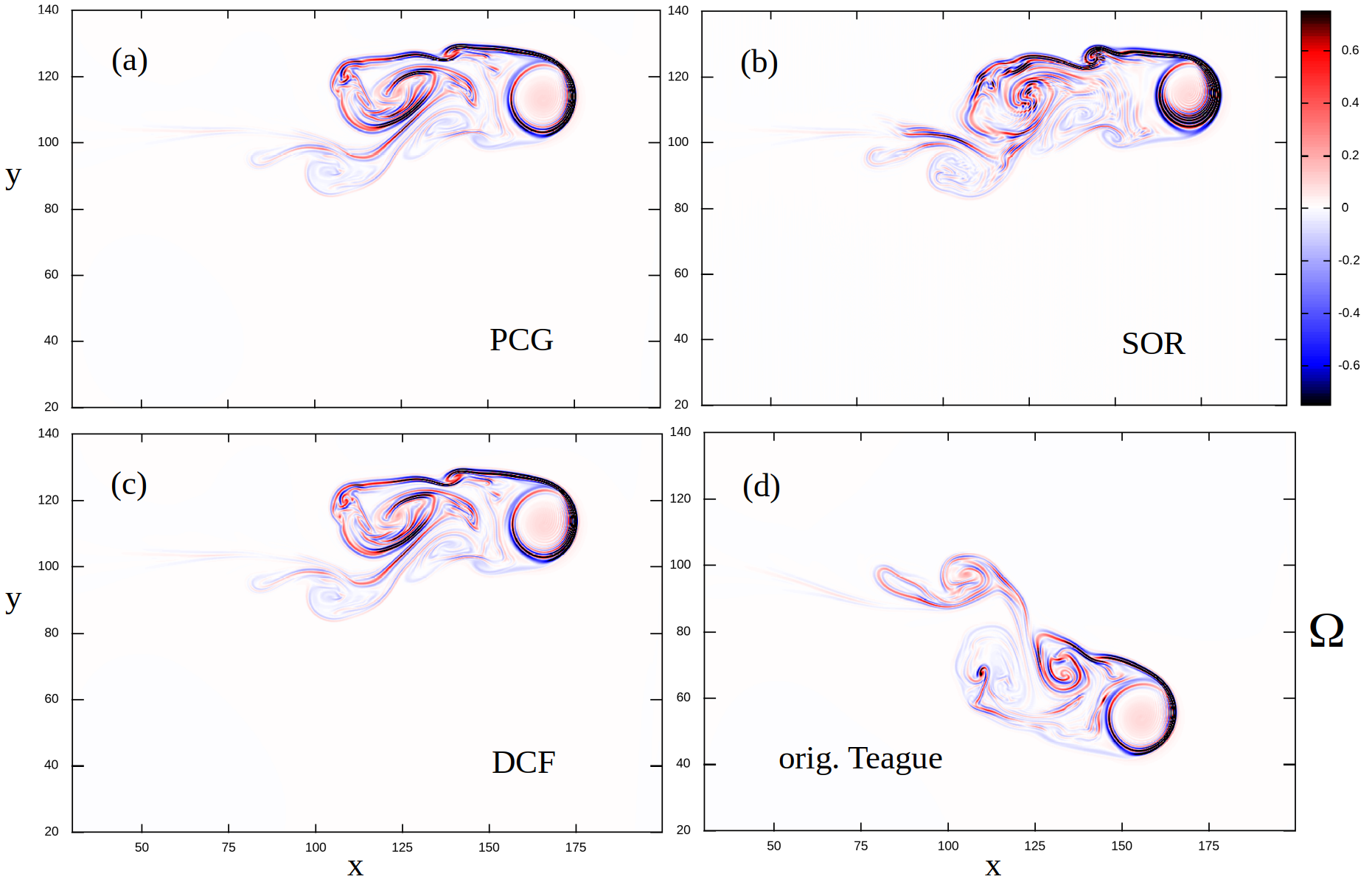}  
\caption{Comparison of vorticity $\hat \Omega = \hat \bnabla^2 \hat \phi(x,y)$
  for an evolved seeded blob ($\tau_i=4$, $\hat N_0=1$, $w=5$) initialised
  with $\hat N_e = \hat N_i$ condition, at $\hat t=2304$, for different generalized
  Poisson solvers. (a) PCG; (b) SOR; (c) DCF; (d) uncorrected (original) Teague solver.} 
\label{f:blob-compare-withv}
\end{figure}

The ExB vorticity $\hat \Omega (x,y)= \hat \bnabla^2 \hat \phi$ of the evolved
blob at the final time $\hat t = 2304$ in units of $\rho_s/c_s$ is compared in
Fig.~\ref{f:blob-compare-vorfree} for simulations with the different PCG, DCF
and SOR solvers used for the polarization. For this plot the zero vorticity
initial conditions is chosen. The blob gradually acquires gyro-induced
FLR spinning and the propagation has characteristic up-down asymmetry.

The three solvers show visually very similar results regarding the
fine structure and global quantities such as the center-of-mass position of
the blob. For comparison the corresponding result obtained with the original
Teague method (without correction) is shown in frame (d). Shape and position
are similar but with clear differences in details. These plots can be compared
with the result of FELTOR simulations of Figure 4 (bottom middle frame) in
ref.~\cite{Held23}. Note that the FELTOR blob has initial position $x_0^{FELTOR} = 50$,
whereas here it is at  $x_0^{TIFF} = 25$. A blob front position of $x_F^{TIFF}  \approx
125$ in our results thus corresponds to a position  $x_F^{FELTOR}  \approx 150$ in the
referenced results and figure.
The overall structure and position is similar, but fine details (e.g.~the tilt of
the blob head, and structures in the secondary trailing vortices) are clearly
different between the TIFF and FELTOR codes, which employ different solvers
for the Poisson problem and the gyro operators. 

Fig.~\ref{f:blob-compare-withv} shows basically the same set-up but with
nonzero vorticity initialisation achieved by setting $\hat N_i = \hat
N_e$. The initial spin largely compensates the later FLR spin build up, so
that propagation has much straighter radial (``to the right'') direction with
less up-down asymmetry.  Again the PCG, DCF and SOR solvers show very close
agreement. Here the result obtained with the original (uncorrected) Teague
method is clearly off and shows a pronounced downward drift. These plots can
again be compared with the corresponding Figure 4 (bottom right) in
ref.~\cite{Held23}. The overall structure is again similar, and details in the
vortex trail again differ. The most noteable difference is in the blob front
position, where $x_F^{TIFF}  \approx 175$ in our present results would be
expected to correspond in FELTOR to $200$, which is already the position of
the right boundary. 
The actual FELTOR blob front position in ref.~\cite{Held23} at the same time
appears to be at $x_F^{FELTOR} \approx 190$ only. A probable explanation for
this difference between the FELTOR and TIFF code results could be in the
respective handling of the (in this particular case very close) boundary conditions. 
The resolutions also differ between the code results, because in
ref.~\cite{Held23} FELTOR uses a discontinuous Galerkin method with 300 grid
cells and 5 polynomial coefficients, which would correspond to 1500 grid
points for the TIFF solvers.
Overall the agreement between the codes and between the different solvers used
in TIFF can be regarded as satisfactory. In particular, the new DCF method has
been shown to achieve strong agreement with the other solvers, and in
particular with the also fourth order PCG scheme.

The results shown in Fig.~\ref{f:blob-compare-withv} have been obtained on a
Threadripper PRO 5975WX Linux workstation with 256 GB RAM, running on 32
threads of the CPU.
The run time was 216~min with the SOR solver, also 216~min for the PCG solver,
67~min for the DCF solver, and 60~min for the uncorrected Teague solver.
The times for the zero vorticity runs (Fig.~\ref{f:blob-compare-vorfree}) were
212~min for the SOR solver, 116~min for the PCG solver, and 68~min for the
DCF solver, and 59~min for the uncorrected Teague solver.
The new DCF solver clearly is most efficient and independent of the complexity
of the density fields. The additional cost for the DCF correction compared to
the original Teague method is around 10-15~$\%$. 

Shorter run times could of course be achieved for the iterative solvers by
reducing accuracy through an increased error tolerance setting.
The resolution with $1024^2$ grid points could be reduced for
practical application scenarios, which would also (for reasons of CFL
stability) allow respectively larger time steps: half of the grid points per
dimension in 2d thus means roughly $1/8$ run time. 
For specific application to study  blob dynamics on a homogeneous
background (such as in the verification example above) the code could also be
sped up by using only one, physical domain with periodic boundary conditions,
instead of the more general quarter-wave extended four-fold mirror
domain. This would reduce computation times by another factor of around 4.

\section{Drift wave turbulence simulations in full-f and full-k}

The paradigmatic Hasegawa-Wakatani (HW) quasi-2d drift wave turbulence fluid
model is in the following extended, as described above, to full-f full-k
isothermal gyrofluid simulations. 
The intention here is again on cross-verification between the polarization
solvers. The 2d polarization and gyro operator solvers can be directly
implemented in future 3d field-aligned gyrofluid turbulence codes, as the
perpendicular (to the magnetic field) 2d drift plane can be solved in the same
manner as in a 2d code like here. A 3d (isothermal electrostatic) resistive
drift wave turbulence code would basically replace the approximate HW coupling
term by direct solution of the parallel electron and ion velocities through
an additional set of nonlinear advective dynamical equations.

\begin{figure} 
\includegraphics[width=8cm]{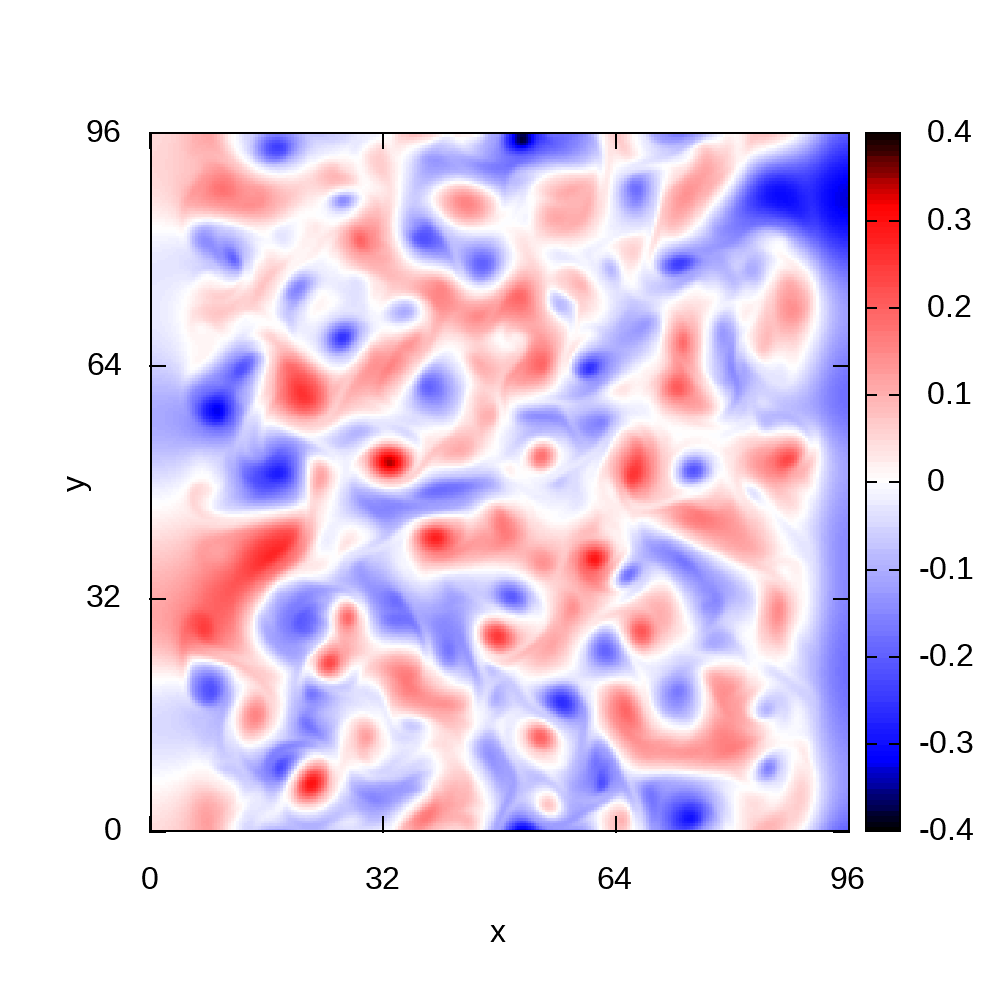}
\includegraphics[width=8cm]{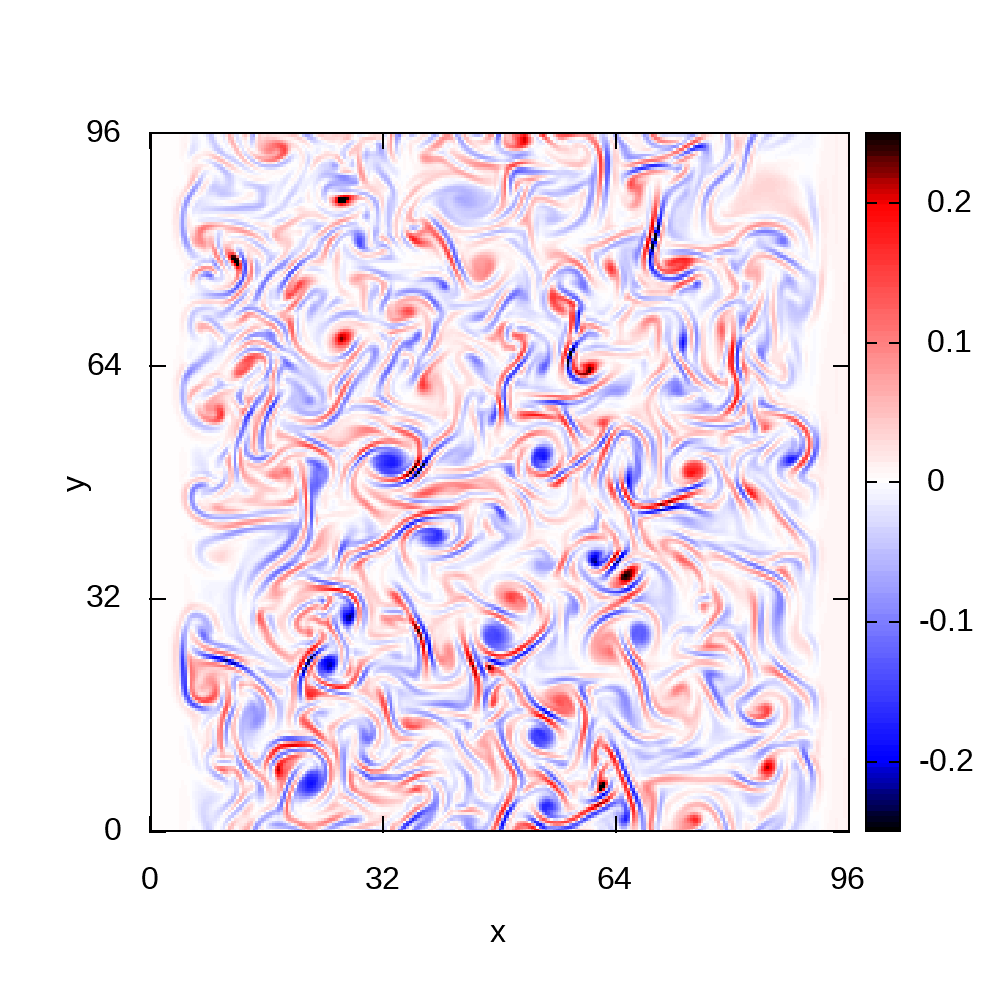}
\caption{Snapshots of (left) potential $\phi(x,y)$ and (right) vorticity
  $\Omega(x,y)$ of drift wave turbulence in the full-f full-k gyrofluid 
  ordinary Hasegawa-Wakatani model, in an $L_x = L_y = 96 \rho_s$ domain.} 
\label{f:ohw-morphology}
\end{figure}

\begin{figure} 
\includegraphics[width=8cm]{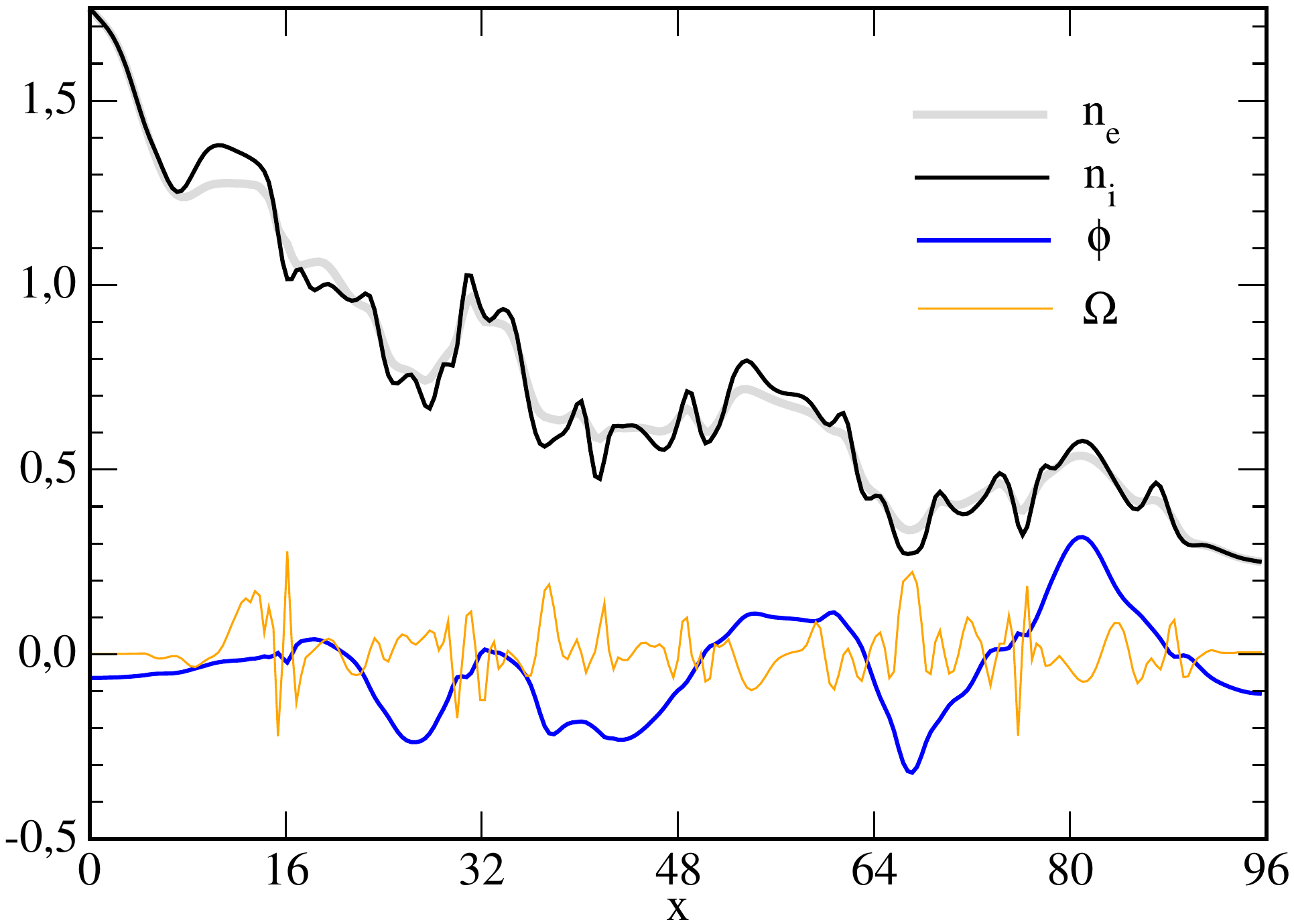}
\caption{Radial cut at $y=L_y/2$ of electron density $n_e(x)$, ion gyrocenter
  density $n_i(x)$, electric potential $\phi(x)$ and vorticity $\Omega(x)$ in
  the saturated phase of OHW turbulence.} 
\label{f:ohw-xcut}
\end{figure}

The morphology of 2d drift wave turbulence in the full-f gyrofluid OHW model
is illustrated in Fig.~\ref{f:ohw-morphology} as a snapshot in the saturated
turbulent phase, for parameters described below. The left frame shows the
electric potential $\phi(x,y)$, which acts as a streamfunction for the
advecting turbulent E-cross-B velocity. The right frame shows the vorticity
$\Omega(x,y) = \bnabla^2 \phi$ with the characteristic thin vorticity sheaths
induced by FLR spin-up \cite{Kendl18}.

Fig.~\ref{f:ohw-xcut} shows a radial cut at $y=L_y/2$ of the electron density
$n_e(x)$, ion gyrocenter density $n_i(x)$, electric potential $\phi(x)$ and
vorticity $\Omega(x)$ at a snapshot in time during the saturated turbulent phase.
Radial boundary conditions are set to zero vorticity and zero zonal flow. The
densities are pinned to the initial background profiles values at the radial
boundaries.

\subsection{Comparison between SOR, PCG and DCF solvers}

The different implemented TIFF solvers are compared for an OHW turbulence case with
$\hat \alpha =0.2$, $\tau_i=1$, $\hat \alpha = 0.2$, $\kappa = 0$, $\delta = 0.015$,
$\hat N_L = 1.75$, $\hat N_R = 0.25$, $L_x=L_y = 96$, $n_x = n_y = 256$,
$\nu_4 = 0.01$, $\Delta t = 0.0025$, and $\beta_x=2$. Initialisation is done
with a random bath of amplitude $\Delta N = 0.05$.

\begin{figure} 
\includegraphics[width=7.2cm]{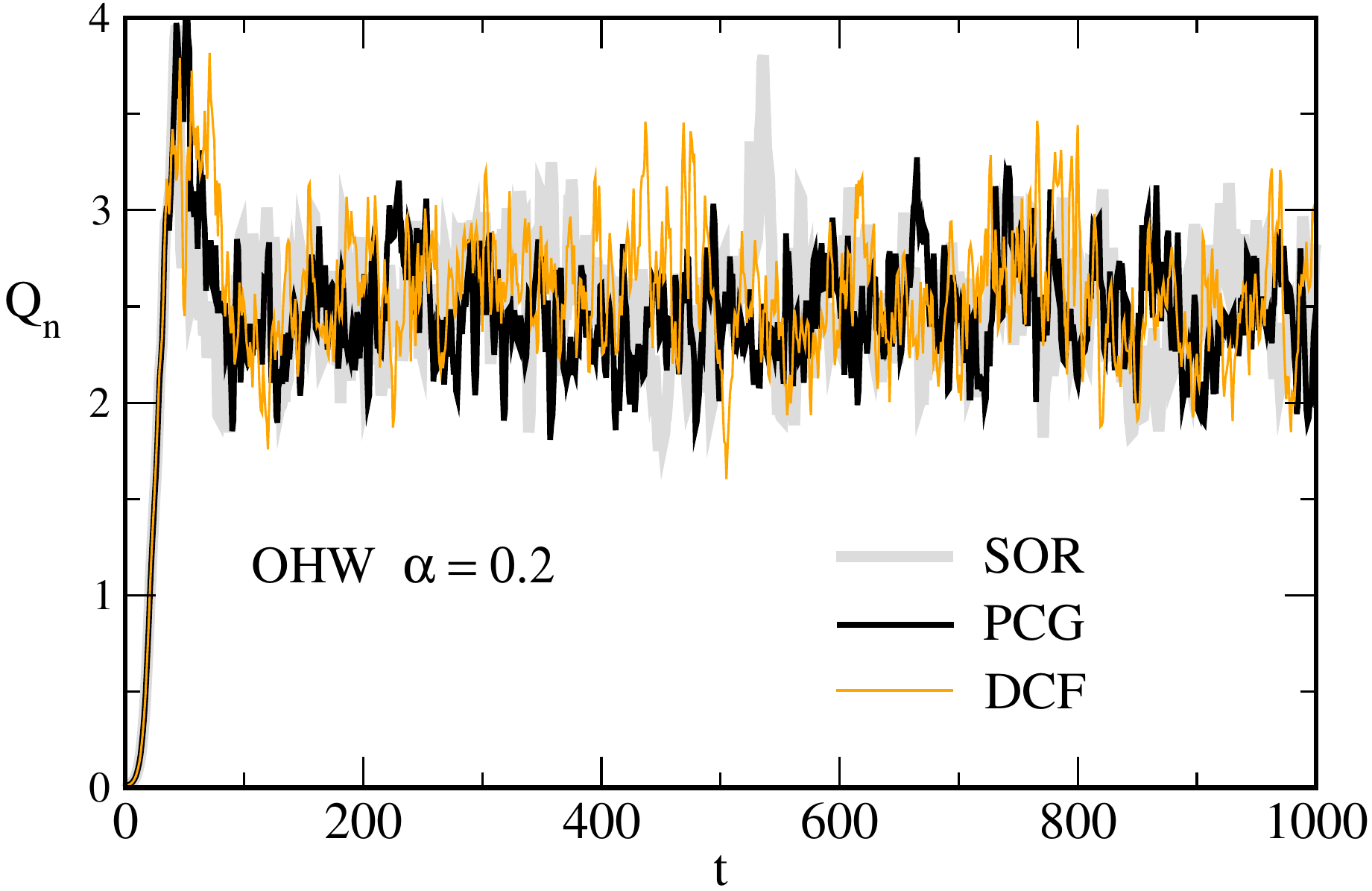}
\includegraphics[width=7.2cm]{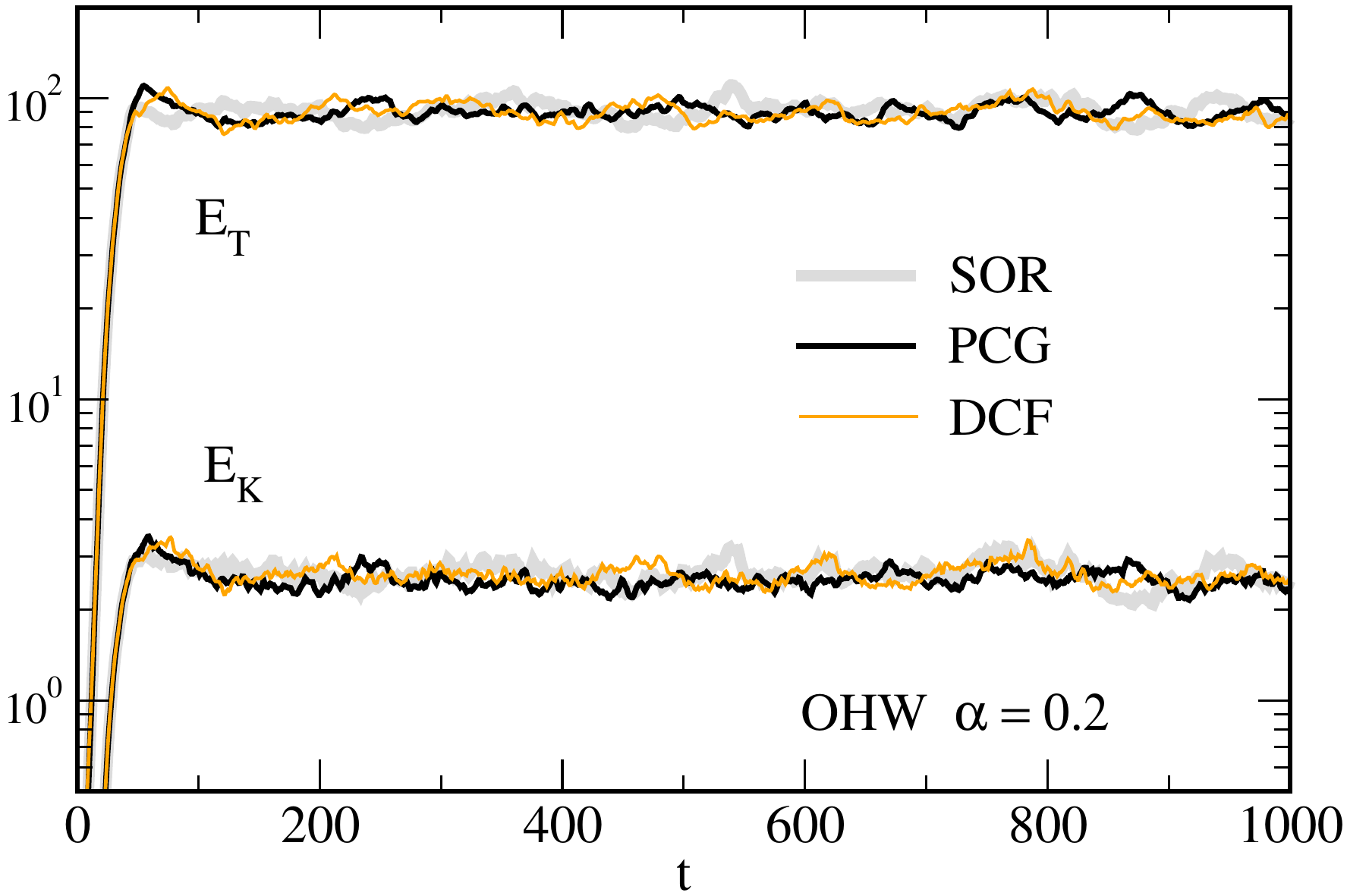}
\caption{Comparison of properties for the SOR, PCG and  DCF solvers. Left:
  transport $Q_n(t)$. Right: thermal energy $E_T(t)$ and   kinetic energy
  $E_K(t)$. Turbulent statistical results (averages and deviations) 
  agree closely for all solvers, and accurately in the initial transient
  quasi-linear phase.} 
\label{f:ohw-solvers}
\end{figure}

Fig.~\ref{f:ohw-solvers} shows in the left frame the turbulent transport
$Q_n(t)$ for the SOR (thick grey), PCG (medium black) and DCF (thin orange)
solvers, and in the right the corresponding time traces of thermal energy
$E_T(t)$ and kinetic energy $E_K(t)$. All time traces agree closely between
the solvers, both in the initial quasi-linear phase and statistically in the
turbulent phase.
The averages in the time window between $100 \leq t \leq 1000$ are:
$\langle Q_n \rangle = 2.50 \pm 0.29$ (SOR), $2.46 \pm 0.26$ (PCG), $2.55 \pm 0.31$ (DCF);
$\langle E_K \rangle = 2.63 \pm 0.21$ (SOR), $2.50 \pm 0.15$ (PCG), $2.61 \pm 0.18$ (DCF);
$\langle E_T \rangle = 91.3 \pm 6.2$ (SOR), $89.6 \pm 5.0$ (PCG), $89.4 \pm 6.1$ (DCF).
The (second order accurate) SOR value of the kinetic energy is around $5$~$\%$ larger compared to
the (fourth order accurate) PCG or DCF results, while the fluctuation
standard deviation of energies for all solvers is also in the order of
$5$~$\%$. The average values of transport and thermal energy agree very well
within the standard deviation for all three solvers.

The computation times were 201 min (SOR), 134 min (PCG) and 81 min (DCF), here
achieved on 16 threads of a a dual Xeon Haswell E5-2687W-v3 workstation.
The (second order) SOR scheme clearly looses in terms of both performance and
accuracy compared to the (fourth order) PCG and DCF schemes, but still can
have its use for testing purposes as a reference generalized Poisson solver
without need for envoking an FFT library. 

The new DCF scheme appears to be both sufficiently accurate and efficient
to be considered for further use.
It should be noted that the 2d FFT evaluation is a memory bound application
and profits from high available bandwidth. The relative performance between
the solvers can therefore differ between hardware systems.

\subsection{Mass and energy conservation test with DCF scheme}

Total mass and total energy should always be ideally conserved by the numerical
scheme employed for the dynamical turbulence simulation.
The change in time of total energy $E(t) = E_K + E_T$ as the sum of the global thermal free
energy and the global kinetic energy (see section \ref{sec:diagnose}) 
can be computed as the nonlinear growth rate $R_E = (1/2E)(\Delta E /
\Delta t)$, and similar the relative rate of change of particle number (or
``mass'') is obtained as $R_M = (1/M)(\Delta M / \Delta t)$ for the total
gyrocenter density as $M(t) = (1/2) [ (\hat N_e(t) - \hat N_0) +  (\hat N_i(t) - \hat N_0)]$.

\begin{figure} 
\includegraphics[width=8cm]{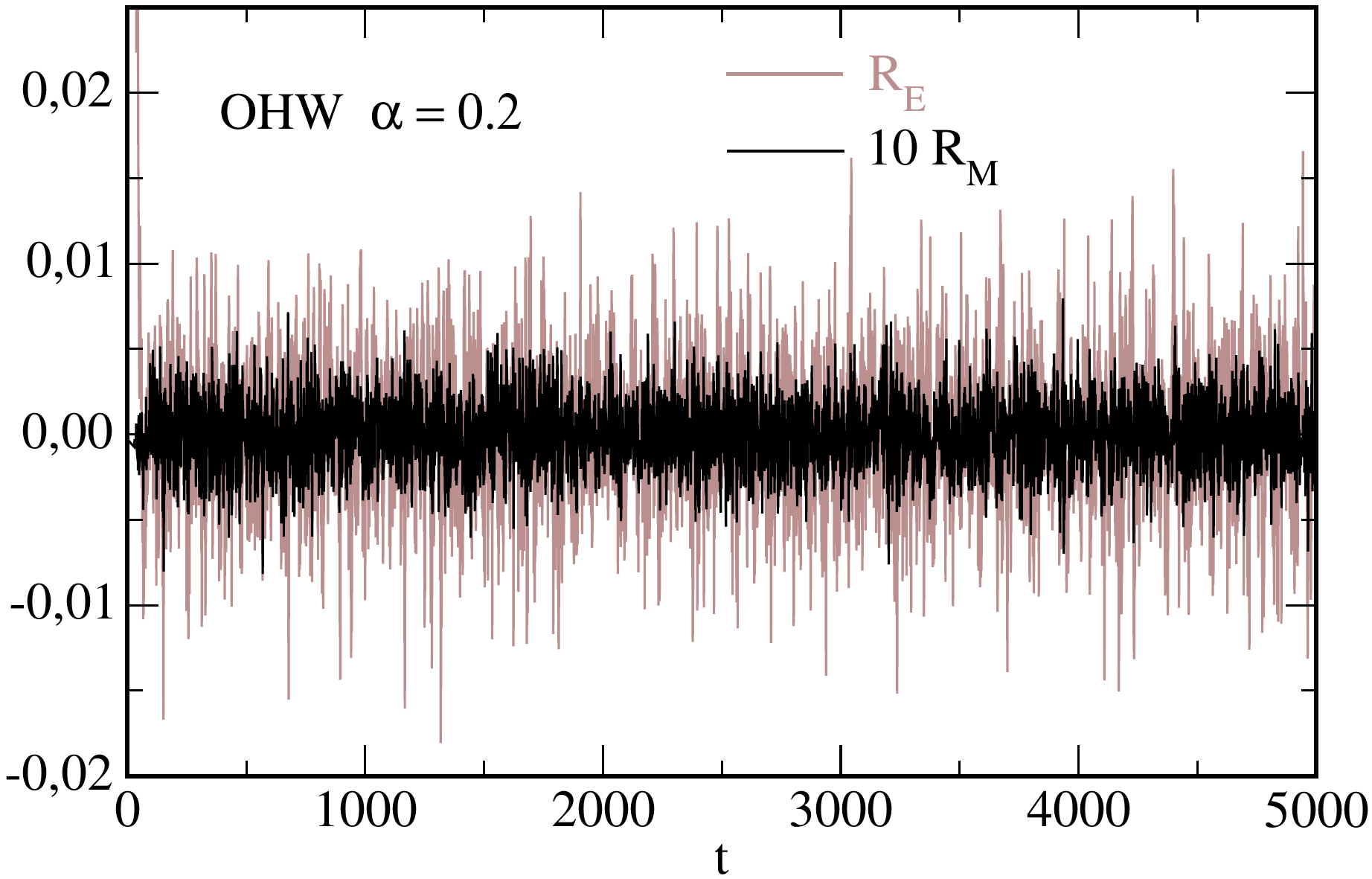}
\caption{
  Conservation of particle number and energy: total energy change rate
  $R_E = (1/2E)(\Delta E / \Delta t)$ and (ten times magnified) gyrocenter density change rate $R_M
  = (1/M)(\Delta M / \Delta t)$ as a function of normalized simulation time.
} 
\label{f:ohwconserve}
\end{figure}

These relative energy and mass change rates as a function of time are shown in
Fig.~\ref{f:ohwconserve} for the same turbulence simulation parameters as
above, obtained with the DCF scheme. A linear regression for the values in
$1000 \leq t \leq 5000$ gives a relative tendency
$\langle R_E \rangle (t) \sim 4 \cdot 10^{-6} - 2 \cdot 10^{-10} \; t$ for the
energy, and $\langle R_M \rangle (t) \sim \cdot 10^{-8} - 4 \cdot
10^{-10} \; t$ for the particle number. Both constitute very small loss rates
and can be regarded as sufficiently good conservation property of the
numerical scheme. The standard deviations of the nonlinear growth rate
fluctuations are $s(R_E) = 4 \cdot 10^{-3}$ and $s(R_M) = 2 \cdot 10^{-4}$.

\subsection{Full-f model in small amplitude limit vs. delta-f model}

The full-f full-k model should agree in the limit of small amplitudes with the
delta-f model, as discussed in sections \ref{transition} and \ref{deltaf}.
To test this we apply largely the same parameters as above to OHW simulations with the
DCF solver: $\hat \alpha = 0.2$, $\tau_i =1$, $\kappa=0$, $\nu_4 = 0.01$, $L_x = L_y =
96$, $n_x=n_y=256$, and $\Delta t = 0.0025$. The inner boundary density is $\hat N_L = 1
+ \epsilon \cdot 0.75$ and outer boundary density $\hat N_R = 1 - \epsilon
\cdot 0.75$, with drift scale $\delta = \rho_s / L_n = \epsilon \cdot
0.015$. A localized ``blob''-like Gaussian perturbation with width $w = 8$ and
amplitude $\Delta \hat N = \epsilon \cdot 0.1$ is initialised.
This parameter set is once run with the delta-f model of section
\ref{deltaf} for $\epsilon = 1$, and once with the complete full-f model but
for $\epsilon = 10^{-3}$ in the small amplitude limit.

\begin{figure} 
\includegraphics[width=8cm]{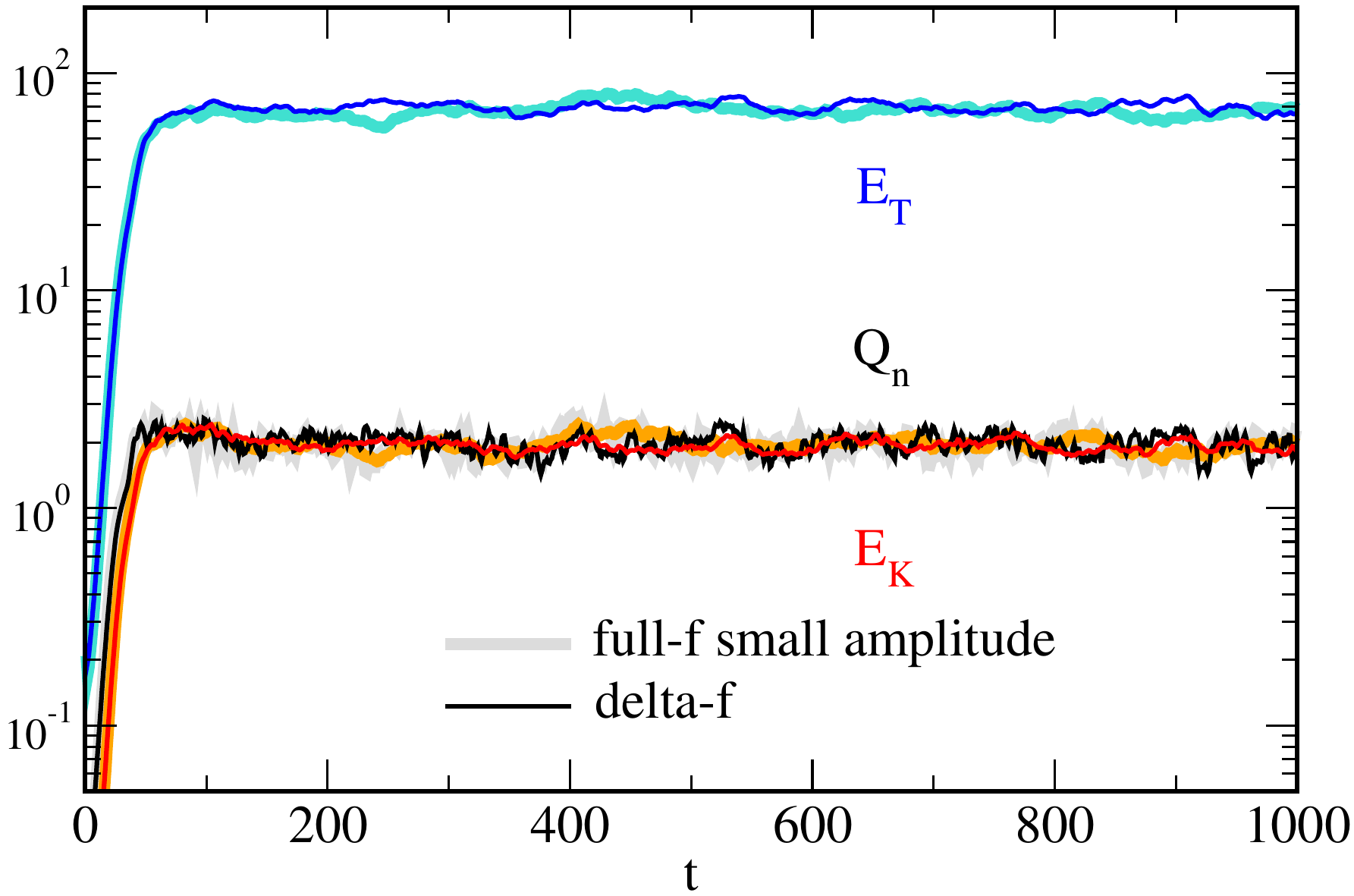}
\caption{Comparison of the full-f full-k model in the small amplitude limit
  with the corresponding delta-f model, for an OHW ($\hat \alpha = 0.2$)
  turbulence simulation. The transport $Q_n(t)$, thermal energy $E_T(t)$ and
  kinetic energy $E_K(t)$ agree closely, both in the quasi-linear transient
  growth phase, and statistically in the nonlinear turbulent phase after around $t>150$.} 
\label{f:ohwdeltaf}
\end{figure}

In Fig.~\ref{f:ohwdeltaf} the time traces of $Q_n(t)$, $E_T(t)$ and $E_K(t)$
are compared for both cases. They show excellent agreement both directly in the
quasi-linear growth phase, and statistically in the nonlinear turbulent phase
after around $t>150$.
Averages in the time window between $200 \leq t \leq 1000$ are:
$\langle Q_n \rangle = 1.95 \pm 0.21$ (FF-$\epsilon$) vs. $1.97 \pm 0.20$ ($\delta$f); 
$\langle E_K \rangle = 1.96 \pm 0.15$ (FF-$\epsilon$) vs. $1.93 \pm 0.11$ ($\delta$f); 
$\langle E_T \rangle = 67.6 \pm 4.7$ (FF-$\epsilon$) vs. $69.6 \pm 3.6$ ($\delta$f).

This demonstrates the correct transition behaviour of the full-f
full-k polarization equation and the dynamical equations towards the delta-f limit. 
The computation time here was about twice as long (81 min) for the full-f
simulation than for the delta-f case (41 min) on the Haswell. The computational expense is
in the evaluation of the Fourier solvers (in the polarization and for the gyro
operators), for which there are eight calls in the full-f DCF case, and four in the delta-f case.

\section{Conclusions and outlook}

An isothermal quasi-2d gyrofluid model and a code (TIFF) for
arbitrary amplitude (full-f) and arbitrary wavelength (full-k) drift
instabilities and turbulence in magnetized plasmas has been introduced.
A major aspect was on testing of a here newly suggested ``dynamically
corrected Fourier'' (DCF) solver for the generalized Poisson equation, which
is based on the original approximate Teague method. The generalized
(a.k.a. ``variable'') Poisson problem appears in the solution of the gyrofluid
(similar to the gyrokinetic) polarization equation, which couples the evolving
gyrocenter plasma densities with the electric potential.
The present approach retains the complete FLR effects and spatial and
dynamical variations of the polarization density. The new DCF solver was shown to
be a viable, sufficiently accurate and efficient method for dynamical
turbulence simulations coupled to the generalized Poisson problem,
which motivates its further use in future extensions of the code.

In its recursive ``RCF'' form (with for example four iteration) as a high
accuracy extension of Teague's method the scheme could also find applications
in the field of optics as a stand-alone application for solution of the
transport of intensity equation (TIE). 

The main purpose of the present work is as reference and test case for the
TIFF model and code, whereas detailed physics studies with the present code,
such as for example on FLR effects on zonal flows, or on properties of
turbulently generated ``blobby'' (intermittent) transport, will be discussed elsewhere.

The extension of the 2d TIFF code to a magnetic field-aligned flux-tube like
3d geometry can directly make use of the here tested polarization solvers, as
these always only act in the locally perpendicular 2d drift plane. Future
developments of such a corresponding 3d full-f full-k gyrofluid edge turbulence code will
include the addition of temperature evolution equations (and through this
applicability on temperature gradient driven modes and use of Landau damping
mechanisms), and generalization to electromagnetic drift-Alfv\'en dynamics.

\section*{Acknowledgment}

The author thanks Markus Held (UiT / UIBK) and Matthias
Wiesenberger (DTU) for valuable discussions and collaboration. 

\section*{Funding}

This work was supported by the Austrian Science Fund (FWF) project P33369.

\section*{Data availability}

The code TIFF is openly available at: https://git.uibk.ac.at/c7441036/tiff


\section*{Appendix A: PCG scheme}

Conjugate gradient methods for the solution of linear systems are
introduced in numerical textbooks such as by LeVeque \cite{leveque}. Here a
pre-conditioned conjugate gradient (PCG) scheme, with a pre-conditioner and
algorithm suggested by Fisicaro et al.~\cite{fisicaro16}, is applied. 

A generalized Poisson operator ${\cal A} \equiv \bnabla \cdot \varepsilon
\bnabla$ is defined so that a solution of the system ${\cal A} \phi = \sigma$ is sought.
The residual of an approximate solution  $\phi^{(n)}$, say obtained in step
$n$ during an iteration, is $r^{(n)} = \sigma - {\cal A} \phi^{(n)}$.
Within the context of a dynamical simulation with small time steps $\Delta t$,
an initial guess can be obtained by extrapolating the converged solutions from
previous time steps towards
\begin{equation}
\phi^{(0)} (t) = \phi (t-1) + a \cdot [ \phi (t-1) - \phi (t-2) ],
\end{equation}
with a free estimation factor $a \in (0, 1)$, and $\phi^{(0)} (t=0) = 0$ in
the first time step. From this the initial residual $r^{(0)}= \sigma - {\cal
  A} \phi^{(0)}$ for the iteration is obtained.

The PCG scheme basically searches for the minimum of a (quadratic) residual function
$f(\phi) = (1/2) (\phi, {\cal A} \phi) -  (\phi,  \sigma)$, which is given
from a 2d Hessian Taylor expansion, by evaluating its (conjugate) gradients
$\bnabla f(\phi) = r = \sigma - {\cal A} \phi $. 
Here $(A,B) = A^T B = \sum_{i,j} A_{i,j} B_{i,j}$ denotes the inner product of
two matrices $A$ and $B$.
A pre-conditioning matrix $q^{(0)}$ is evaluated once (in each dynamical 
time step) and remains constant throughout the PCG iteration.

The iteration algorithm (cf.~sec.~5.3.5 in ref.~\cite{leveque}, and table 2 in
ref.~\cite{fisicaro16}) proceeds as: 

\noindent (0) Pre-process $q^{(0)} \equiv \sqrt{\varepsilon} \bnabla^2 \sqrt{\varepsilon}$.

\noindent (1) $v^{(n)}  \equiv {\cal P}^{-1} (r^{(n)})$, with a precondition operator
${\cal P}$ specified below.

\noindent (2) $\beta^{(n)} = (v^{(n)}, r^{(n)}) / (v^{(n-1)},
r^{(n-1)})$, for $n \neq 0$.

\noindent (3) $p^{(n)} = v^{(n)} + \beta^{(n)} p^{(n-1)}$.

\noindent (4) $w^{(n)} = {\cal A} p^{(n)} =  {\cal A} v^{(n)} +  \beta^{(n)} {\cal
  A} p^{(n-1)} =  r^{(n)}  - q^{(0)} v^{(n)} +  \beta^{(n)}  w^{(n-1)}$.

\noindent (5) $\alpha^{(n)} = (v^{(n)}, r^{(n)}) / (p^{(n)}, w^{(n)})$.

\noindent (6) $\phi^{(n+1)} = \phi^{(n)} + \alpha^{(n)} p^{(n)}$.

\noindent (7) $r^{(n+1)} = r^{(n)} - \alpha^{(n)} w^{(n)}$.

\noindent (8) Return to step (1) until the residual
$||r^{(n+1)}||$ is below a specified limit. 

The algorithm allows to efficiently re-use several already calculated terms
and products.
A classical conjugate gradient (CG) solver without preconditioning of
the residual in step (1) would set ${\cal P} = 1$ and $v^{(n)} = r^{(n)}$ in the algorithm.
The idea is that a preconditioned residual $v = {\cal P}^{-1} r = {\cal
  P}^{-1}\sigma - {\cal P}^{-1}{\cal A} \phi $ is minimized in the same way
as the original $r$. 

Application of a suitable preconditioner can speed up convergence
significantly, but needs to be chosen carefully. Here the preconditioner
suggested in ref.~\cite{fisicaro16} for the generalized Poisson problem is
applied as ${\cal P} () \equiv \sqrt{\varepsilon} \bnabla^2 \sqrt{\varepsilon}
()$, and accordingly $q^{(0)} \equiv \sqrt{\varepsilon} \bnabla^2
\sqrt{\varepsilon}$ above.

The evaluation of ${\cal A} v = \bnabla \cdot \varepsilon \bnabla v =
\sqrt{\varepsilon} \bnabla^2 v \sqrt{\varepsilon} - v \sqrt{\varepsilon} 
\bnabla^2 \sqrt{\varepsilon} = {\cal P} (v) - v q^{(0)} = r - v q^{(0)}$ in step
(4) is greatly simplified by this preconditioner \cite{fisicaro16}.
For the inversion in step (1) here $v^{(n)} = {\cal
  P}^{-1} (r^{(n)}) = (1/\sqrt{\varepsilon}) \bnabla^{-2} (  r^{(n)} /\sqrt{\varepsilon} )$
needs to be evaluated including the application of a standard fast Poisson solver,
which is here presently achieved by an FFT solver in ${\bf k}$ space.
The additional expense of calling a standard Poisson ($\bnabla^2 u = f$)
solver at each iteration step is paid off by the in general rapid convergence
of this PCG scheme. The default choice for the standard Poisson solver in the
TIFF code is by FFT, but optionally also a (generally slower) SOR scheme is available.
The order of this PCG solver is determined by the order of evaluation of the
Laplacians (in $q^{(0)}$, or in a non-FFT standard Poisson solver), which is
here achieved in fourth order accuracy.

The algorithm above could also be converted into a (preconditioned) steppest
descent scheme by setting $\beta^{(n)} = 0$, which more serves didactical than
practical purposes because of its generally slower convergence. 


\section*{Appendix B: SOR scheme}

Successive over-relaxation (SOR) is an established method for 
iterative solution of elliptic equations, such as the generalized Poisson
equation $\bnabla \cdot \varepsilon \bnabla \phi = \sigma$ of electrostatics,
through an appropriate finite difference discretization. Here a basic
algorithm and second order discretization as outlined in the textbooks by
LeVeque \cite{leveque} and Humphries \cite{humphries} is followed.  

In one dimension, the inner term $\varepsilon \partial_x \phi \approx W \;
\Delta \phi / \Delta x$ is first discretized with interpolated coefficients $W =
(\varepsilon_i + \varepsilon_{i-1})/2$, and $ \Delta \phi = \phi_i -
\phi_{i-1}$. In the following an equidistant rectangular grid with 
$\Delta x = \Delta y \equiv h$ is assumed. Applying likewise the outer
derivative, one gets $W_{i+1} (\phi_{i+1} - \phi_i) - W_{i-1} (\phi_i -
\phi_{i-1})  = h^2 \sigma$, with coefficients $W_{i+1} = (\varepsilon_{i+1} +
\varepsilon_i)/2$ and $W_{i-1} = (\varepsilon_{i} + \varepsilon_{i-1})/2$,
which can be re-arranged to define $\phi_i$ at grid node $x_i$.

In the same manner, the 2d relation for $\phi_{i,j}$ in 2nd order 5pt-stencil is obtained as
\begin{eqnarray}
\phi_{i,j} &=& \frac{1}{W_0} \left( W_{i+1,j} \phi_{i+1,j} +   W_{i-1,j}
               \phi_{i-1,j} +  \right. \nonumber \\
           & &  \left.  +   W_{i,j+1} \phi_{i,j+1} +   W_{i,j-1} \phi_{i,j-1} -   h^2
  \sigma_{i,j}  \right)
\label{eq:phiij}
\end{eqnarray}
with $W_0 =  W_{i+1,j} +   W_{i-1,j} +   W_{i,j+1}  +   W_{i,j-1}$, and
$W_{i+1,j}  = (\varepsilon_{i+1,j} + \varepsilon_{i,j})/2$,
$W_{i-1,j}  = (\varepsilon_{i-1,j} + \varepsilon_{i,j})/2$,
$W_{i,j+1}  = (\varepsilon_{i,j+1} + \varepsilon_{i,j})/2$,
$W_{i,j-1}  = (\varepsilon_{i,j-1} + \varepsilon_{i,j})/2$.


Starting with an initial guess $\phi^0({\bf x})$, the relation (\ref{eq:phiij})
can be iterated, with grid values of $\phi_{i,j}^{(n)}$ applied on the r.h.s. in
order compute the updated  $\phi_{i,j}^{(n+1)}$ on the left, until the
residual error norm
$|| R^{(n+1)} ||$ with $R^{(n+1)} = \phi^{(n+1)} - \phi^{(n)} $ is smaller than a specified limit.

The SOR method \cite{leveque} improves convergence by applying a correction factor $\omega$ as
\begin{equation}
\phi^{(n+1)} = \phi^{(n)} + \omega R^{(n)}.
\end{equation}
The grid array is swept in odd (``red'') - even (``black'') order, as in
eq.~(\ref{eq:phiij}) the values of $\phi_{i,j}^{(n)}$ for even indices $i$ or
$j$ depend only on odd indexed grid values of  $\phi^{(n-1)}$, and vice versa.
The over-relaxation parameter $\omega$ is determined by Chebyshev acceleration
according to $\omega_{odd}^0 = 1$ and  $\omega_{even}^0 = 1/(1-r^2/2)$ as
initial values, and further $\omega = 1/(1-r^2 \omega/4)$, updated in each half-sweep.
Here the spectral radius is calculated as $r = [ \cos(\pi / n_x) + \cos(\pi /
n_y)]/2$ for a homogeneous equidistant grid with grid point numbers $n_x$ and $n_y$.
Each ``red'' and ``black'' sweep through the 2d checkerboard-like grid is loop
parallelized with OpenMP, respectively. 

Convergence can be significantly accelerated by an initial guess for $\phi^0({\bf
  x})$ based on the solution from previous time step(s) within a dynamical
simulation for small time steps $\Delta t$, by extrapolating (in similar
spirit as for the over-relaxation factor above)
\begin{equation}
\phi^{0} \equiv \phi(t-1) + a \cdot (\phi(t-1) - \phi(t-2) ).
\end{equation}
with a free prediction factor $a \in (0, 1)$. The first time step uses
$\phi^{0} = 0$, in the second step $a=0$, and in later times a factor
between $0.5$ and $1.0$ has been found to be most efficient.

The rate of convergence depends on the degree of (non)uniformity of
$\varepsilon({\bf x})$. In classical electrostatics the permittivity is
usually only weakly varying within one medium, but may have discontinuities
between neighbouring media, and is often set in complicated geometries, so
that mostly rather finite element schemes instead of finite
difference schemes are employed. In the present application on the gyrofluid 
polarization equation in a small local section of an edge plasma, the grid can
be simply chosen as rectangular, but the polarization density can be strongly inhomogeneous for
large amplitude turbulent fluctuations.

The present SOR implementation is only second order accurate, but straightforward
to implement and can serve as a reference for the fourth order PCG and DCF
schemes.
For consistency, all dynamical simulations shown in this publication that have
been obtained with the second order SOR scheme also use the second order
versions of the Arakawa scheme (for the advecting Poisson brackets) and of the
curvature operator. For all other generalized Poisson sovers (DCF, PCG) the
respective consistent fourth order versions are used.

For evaluation of the standard Poisson problem ($\bnabla^2 u =  f$) also a
fourth order accurate SOR scheme is available in the TIFF code, which uses
discretization by a Collatz Mehrstellenverfahren (9pt stencil for the Laplacian
on $u$, and a 5pt stencil correction for $f$).


\section*{Appendix C: DCF scheme}

The here introduced ``dynamically corrected Fourier'' (DCF) approach for the
generalized Poisson problem adds a correction term, computed from the result
of the previous time step within a dynamical simulation, to ``Teague's method''
(compare main text above).  
Teague's approximate method \cite{teague83} had originally been devised for
solution of the TIE (transport of intensity equation) in optics \cite{zuo20},
and can be efficiently evaluated in Fourier space \cite{paganin98}.

It is here innovatively applied on solution of the electric potential $\phi
({\bf x})$ from the gyrofluid (or gyrokinetic) polarization equation, and 
within a dynamical context. For small time step sizes $\Delta t$ this allows
to re-use solutions of $\phi^{old}$ from past times, instead of a (more
expensive) iterative error correction. An extrapolated prediction with a free
estimation parameter $a \in (0, 1)$ based
on two previous solutions is used for the corrector, as
\begin{equation}
\phi^{old} \equiv \phi^{(t-1)} + a \cdot (\phi^{(t-1)} - \phi^{(t-2)} ).
\end{equation}

The generalized Poisson equation $\bnabla \cdot \varepsilon \bnabla \phi = \sigma$ 
is formulated by means of a polarization density vector field ${\bf P} \equiv \varepsilon
\bnabla \phi \equiv \bnabla p + \bnabla \times {\bf H}$ in terms of a scalar
potential $p$ and a vector potential ${\bf H} = \eta \; {\bf e}_z$.
Input quantities are the known 2d fields $\sigma ({\bf x})$ and  $\varepsilon ({\bf x})$.

By application of the divergence on $\bnabla \cdot {\bf P} = \sigma$
the scalar field $p$ is first obtained from:
\begin{equation}
  p  = \bnabla^{-2} \sigma.
\end{equation}
The inversion of the Laplacian in the standard Poisson problem is here
obtained in ${\bf k}$ space by Fourier transforms, using the FFTW3 library, as
$ p  = - {\cal F}^{-1} k^{-2} {\cal F} \sigma$. 

Further, an approximation for $A$ is obtained (see main text above) from $\bnabla \times {\bf P}$ as:
\begin{equation}
  \eta_o  ({\bf x}) \equiv  \bnabla^{-2} \{ \phi^{old}, \varepsilon \}.
 \label{eq:dcf-ao} 
\end{equation}
This estimated quantity is used in the extended, dynamically corrected relation:
\begin{equation}
  \phi ({\bf x})  = \bnabla^{-2} \left[ \bnabla \cdot \frac{1}{\varepsilon } 
  \bnabla  p + \; \left \{ \frac{1}{\varepsilon}, \eta_o
  \right \} \right].
 \label{eq:dcf-phi} 
\end{equation}
Like above, the inverted Laplacians in eqs.~(\ref{eq:dcf-ao}) and
(\ref{eq:dcf-phi}) can be efficiently solved in Fourier space. The term
$\bnabla \cdot (1/\varepsilon) \bnabla  p$ is evaluated by standard 
(fourth order) centered finite differences, and the Poisson bracket can be
computed either also by simple fourth order centered differences, or by
re-using the Arakawa scheme (introduced for the advective terms in the
dynamical gyrocenter density equations). 

For a solver unit test (with a constructed solution) the set of equations
(\ref{eq:dcf-ao}) and (\ref{eq:dcf-phi}) can also be applied recursively
(instead of within a dynamical context) by re-setting $\phi \rightarrow
\phi^{old}$ directly after each iteration step, starting with $\eta_o = 0$.
In the main text above this is refered to as ``RCF'' for recursively corrected
Fourier scheme. 

The computational bottleneck in this DCF solver for the generalized Poisson
equation is provided by the three forward plus backward 2d
Fourier transforms (in the three $\bnabla^{-2} = - {\cal F}^{-1} k^{-2} {\cal
  F}$ operations).
To put this in context, note that in addition, Fourier solvers are here presently
also used for evaluation of all gyro-operators: 
application of any (PCG, SOR or DCF) solver on the full-k polarization
equation requires two further, here also Fourier based, evaluations of the
$\sqrt{\Gamma_0}^{-1}$ operator (in TIFF code algorithm steps $\#$4 and 5),
and in general another in computing $\hat N_{Gi}$ for use in $\sigma$ in step
$\#$2, one more for the ion potential $\phi_i$, and a further one in the
boundary conditions (step $\#4$). 
In total eight forward plus backward transforms are employed per time step.
The execution time per Fourier solver application depends only on the size of
the grid arrays (and the chosen transform method or library), which by
extrapolation allows predictable run time estimates. 
For cold ion cases ($\tau_i = 0$) without FLR effects the four gyro-operations herein
can be bypassed, and computation time is significantly reduced.

For comparison, the delta-f polarization equation requires only one call of a
Fourier solver routine (regardless of cold or warm ions) for a given $\sigma$,
so the execution time for a full-f full-k polarization solve is more than five
times longer compared to a delta-f solution. (When iterative solvers such as
PCG or SOR are used, the execution time could be tweaked by compromising on accuracy.)


\end{document}